\documentclass[fleqn,usenatbib]{mnras} 
\usepackage{pgfplots}
\pgfplotsset{compat=1.14}

\usepackage[T1]{fontenc}
\usepackage{lmodern}

\usepackage{ae,aecompl}
\usepackage{verbatim}

\usepackage{graphicx}	

\usepackage{amssymb}	
\usepackage{multicol}
\usepackage{multirow}
\usepackage{pdflscape}	
\usepackage{amsmath} 
\usepackage{cleveref}

\usepackage{color}

\usepackage{times}
\usepackage[space]{grffile}
\usepackage[normalem]{ulem}
\usepackage{natbib}
\usepackage{longtable}
\usepackage{hyperref}
\usepackage{comment}
\bibpunct{(}{)}{;}{a}{}{,}
\usepackage{pdflscape}	
\usepackage{url}
\texorpdfstring

\DeclareRobustCommand{\VAN}[3]{#2}
\let\VANthebibliography\thebibliography
\def\thebibliography{\DeclareRobustCommand{\VAN}[3]{##3}\VANthebibliography}

\usepackage{graphicx}	
\usepackage{cite}
\usepackage{orcidlink}

\usepackage{tikz}
\usetikzlibrary{matrix}

\usepackage{graphicx}	
\usepackage{amsmath}	
\usepackage{amssymb}	
\usepackage{multicol}
\usepackage{multirow}
\usepackage{pdflscape}	
\usepackage{amssymb}	

\usepackage{color}
\usepackage{times}
\usepackage[space]{grffile}
\usepackage{cleveref}
\usepackage[normalem]{ulem}
\usepackage{natbib}
\usepackage{longtable}
\usepackage{hyperref}
\usepackage{comment}
\bibpunct{(}{)}{;}{a}{}{,}
\usepackage{pdflscape}	
\usepackage{url}
\texorpdfstring

\usepackage[T1]{fontenc}
\usepackage{ae,aecompl}
\usepackage{verbatim}
\usepackage[varg]{txfonts}

\newcommand{\fek}{Fe~K$\alpha$}
\newcommand{\xmm}{{\em XMM-Newton}}
\newcommand{\nustar}{{\em NuSTAR }}
\newcommand{\chandra}{{\em Chandra}}
\newcommand{\suzaku}{{\em Suzaku}}

\newcommand{\swift}{{\small \it Swift}}

\newcommand{\xrt}{{\small {\it Swift}/XRT}}

\newcommand{\xclumpy}{{\tt xclumpy}}

\newcommand{\uxcl}{{\tt UXCLUMPY}}

\newcommand{\hb}{H$\beta$}

\newcommand{\oiii}{[O{\,\sc iii}]}

\graphicspath{{./}{pic/}}

\title[OXCLAGN]{The {\it NuSTAR} view of five changing-look active galactic nuclei}

\author[Bing Lyu et al.]{
Bing Lyu \orcidlink{0000-0001-8879-368X},$^{1}$\thanks{E-mail:lyubing@pku.edu.cn}
Zhen Yan \orcidlink{0000-0002-5385-9586},$^{2}$
Xue-bing Wu \orcidlink{0000-0002-7350-6913},$^{1,3}$
Qingwen Wu\orcidlink{0000-0003-4773-4987},$^{4}$
Wenfei Yu \orcidlink{0000-0002-3844-9677}, $^{2}$
Hao Liu \orcidlink{0000-0001-5525-0400}$^{5}$
\\
$^{1}$Kavli Institute for Astronomy and Astrophysics, Peking University, Beijing 100871, People's Republic of China\\
$^{2}$Shanghai Astronomical Observatory, Chinese Academy of Sciences, 80 Nandan Road, Shanghai, 200030, People's Republic of China \\
$^{3}$Department of Astronomy, School of Physics, Peking University, Beijing, 100871, People's Republic of China \\
$^{4}$Department of Astronomy, School of Physics, Huazhong University of Science and Technology, 1037 Luoyu Road, Wuhan, 430074, People's Republic of China \\
$^{5}$University of Science and Technology of China, No.96, JinZhai Road Baohe District, Hefei, Anhui, 230026, People's Republic of China \\
}

\pubyear{2024}

\begin{document}
\label{firstpage}
\pagerange{\pageref{firstpage}--\pageref{lastpage}}
\maketitle

\begin{abstract}
Changing-look active galactic nuclei (CLAGNs) are known to change their spectral type between 1 and 2 (changing-state) or change their absorption between Compton-thick and Compton-thin (changing-obscuration) on timescales of years or less. The physical mechanism and possible connection between the two types of CLAGNs are still unclear. We explore the evolution of the broadband X-ray spectra from Nuclear Spectroscopic Telescope Array (\nustar\,) and column density in five CLAGNs with moderate inclination viewing angles, which have shown significant variations of both optical types and X-ray absorption. Based on a phenomenological and two clumpy torus models, we find that the X-ray photon index ($\Gamma$) and the Eddington-scaled X-ray $2-10$ keV luminosity ($L_{\rm X}/L_{\rm Edd}$) are positively correlated for the five sources, which are similar to other bright AGNs and optical CLAGNs at type 1 phase. We find a significant negative correlation between log$N_\mathrm{H,los}$ and log$L_{\rm X}/L_{\rm Edd}$ except for ESO 362-G18.  Similar to changing-state AGNs, changing-obscuration AGNs may be also triggered by the evolution of the accretion disc. Our results support the disc wind scenario, where the disc wind proportional to the accretion rate and formed at moderate inclination angles would push the obscuration material further away and decrease the column density from the line of sight observed in the changing-look AGNs.


\end{abstract}


\section{Introduction}

An Active Galactic Nucleus (AGN) is a compact region at the center of a galaxy, which is powered by the material accreting onto a supermassive black hole \citep[SMBH, e.g.,][]{2008ARA&A..46..475H}. AGNs are empirically divided into type 1 AGNs with both broad emission lines (e.g., $\rm FWHM  \gtrsim 1000\,km/s$) and narrow emission lines ($\rm FWHM \lesssim 500\,km/s$) and type 2 AGNs with only narrow emission lines. The sub-classes (e.g., Seyfert 1.5, 1.8, and 1.9) are further defined with the declines of the broad $\rm H\beta$ emission lines relative to narrow \oiii\, lines  \citep[][]{1976MNRAS.176P..61O,1981ApJ...249..462O}. The broad H$\beta$ lines are undetectable and very weak in type 1.9 and type 1.8 AGN. The broad H$\alpha$ and H$\beta$ lines are comparable in type 1.5 AGN.  The difference between the two types of AGNs is attributed to the different inclination angles with respect to the line of sight in the so-called unified model of AGN \citep[e.g.,][]{1993ARA&A..31..473A,Netzer2015}. In this scenario, the broad emission lines come from the broad line region (BLR) at the sub-pc scale and the narrow emission lines come from the narrow line region (NLR) at the kpc scale. Type 1 AGNs are face-on viewed with the broad emission lines visible to us, while type 2 AGNs are edge-on and the broad emission lines are obscured by the surrounding dusty torus. Their intrinsic properties are considered to be the same and their optical classification and absorption from torus should not change in such a short timescale of years or even months. However, two kinds of CLAGNs, optical (appearance/disappearance of broad emission lines) and X-ray CLAGNs (switch between Compton thin \& Compton-thick state) are observed in recent years, which propose challenges to the traditional unified model of AGN. 

The optical changing-look AGNs are also referred to as changing-state AGN, \citep[CS-AGN; e.g.][]{2023NatAs...7.1282R}, where their broad emission lines appear or disappear within several months to years \citep[e.g.,][]{2014ApJ...796..134D,2014ApJ...788...48S,2015ApJ...800..144L,2018ApJ...862..109Y,2019ApJ...887...15W,2019ApJ...883...94T,2019MNRAS.486..123R,2019MNRAS.487.4057K,2020ApJ...901....1W,2023ApJ...953...61Y}. The traditional explanation is due to the obscuring clouds moving into the line of sight and block the broad emission lines \citep[e.g.][]{2009MNRAS.393L...1R}. However, this explanation can only be applied to a few sources \citep[e.g.][]{2013MNRAS.436.1615M,2014MNRAS.443.2862A,2015ApJ...815...55R,2018MNRAS.481.2470T,2022ApJ...939L..16Z}. Besides, this scenario is inconsistent with the low $N_\mathrm{H}$ and significant mid-infrared (MIR) variability found for most of the optical CLAGNs \citep[e.g.][]{2017ApJ...846L...7S,2018MNRAS.480.3898N}. Another explanation is related to the change of the accretion rate \citep[e.g.][]{2024arXiv241108676J,2024ApJ...966...85Z,2024ApJS..272...13P}. Those optical CLAGNs with suppression/enhancement of the blue continuum and disappearance/appearance of the broad emission lines are usually accompanied by the accretion-rate driven multi-wavelength and rapid variability. On the one hand, as the accretion rate decreases, it is hard to sustain the BLR, and AGN would follow a sequence from type 1, 1.2/1.5,  1.8/1.9, to 2 in a disc-wind BLR model \citep[e.g.][]{2014MNRAS.438.3340E}. On the other hand, the inner part of standard Shakura-Sunyaev thin disc \citep[SSD; ][]{1973A&A....24..337S} could be truncated by a radiatively inefficient accretion flow (RIAF) \citep[e.g.,][]{2008ARA&A..46..475H,2014ARA&A..52..529Y} when the accretion rate is lower than a critical value and the disc provides deficient ionizing photons to excite the gas in the BLR \citep[e.g.][]{2018MNRAS.480.3898N,2021MNRAS.508..144G}.  The 6.4 keV Fe-K$\alpha$ line is a ubiquitous feature in AGNs' X-ray spectra, which could be a probe of the BLR or the torus materials \citep[e.g.,][]{2022arXiv221202731N}. The dust sublimation radius serves as an outer envelope of the Fe-K$\alpha$ line emitting region \citep[e.g.,][]{2015ApJ...812..113G}, which is confirmed in most AGNs with the narrow Fe-K$\alpha$ regions smaller than dusty sublimation radius \citep{2022A&A...664A..46A}.  The narrow Fe-K$\alpha$ line emitting region from the reverberation mapping for NGC 3516 at type 2 phase is consistent with the location of BLR at type 1 phase and is significantly smaller than the dusty torus radius \citep{2022arXiv221202731N}, which suggests the BLR materials remain during the type 2 phase and confirms the latter scenario.

X-ray CLAGNs are originally referred to as AGNs that show variable X-ray spectra switching from/to Compton-thick AGNs ($N_\mathrm{H} \gtrsim \mathrm{10^{24}\,cm^{-2}}$) to/from Compton-thin AGNs ($N_\mathrm{H} < \mathrm{10^{24}\,cm^{-2}}$) \citep[e.g.,][]{2003MNRAS.342..422M,2007MNRAS.377..607P,2009ApJ...695..781B,2014MNRAS.443.2862A,2015ApJ...815...55R,2016ApJ...820....5R,2022ApJ...935..114M}.  The line of sight column density ($N_\mathrm{H}$), which reflects the properties of the torus, can be determined by the X-ray spectral fitting since the obscuring materials significantly absorb the emission below soft X-ray (e.g. $<10 \rm \,keV$) band due to the internal dust extinction \citep[][]{2002RSPTA.360.2045M}. Those AGNs with a rapid and extreme variation of $N_\mathrm{H}$ (e.g., variation larger than one order of magnitude in timescales of years or months) could be also considered as X-ray CLAGNs or changing-obscuration AGN \citep[CO-AGN; see a recent review in][]{2023NatAs...7.1282R}. For the changing-obscuration AGNs, the rapid variation of the column density can be attributed to the motion of obscuring clouds \citep{2016ApJ...820....5R} or the clumpy tours \citep{2009MNRAS.393L...1R}. Otherwise, it might also be related to the change of intrinsic accretion rate, e.g., the change in the ionization state of the obscuring material \citep[e.g.][]{1989MNRAS.236..153Y,2023MNRAS.tmp..710N,2023NatAs...7.1282R}, the sublimation of dust in the clouds near the BLR \citep[e.g.][]{2021MNRAS.507..687J}, or the disc wind associated with the variation of the intrinsic AGN luminosity \citep[e.g.][]{2017Natur.549..488R,2022A&A...662A..77M}.

The number of CLAGNs is growing quickly but the total number is still limited (\citealp[$\lesssim 300$ for optical CLAGNs, e.g.,][]{2019ApJ...874....8M,2020ApJ...889...46S,2021A&A...650A..33P,2021MNRAS.503.2583S,2022ApJ...926..184J} and \citealp[$\sim 23$ for X-ray CLAGNs, e.g.,][and references therein]{2022ApJ...927..227L,2022ApJ...935..114M}). Less than 13 CLAGNs show repeating behavior of appearance/disappearance of broad emission lines \citep[e.g.][]{2020A&A...641A.167S,2022RAA....22a5011W}. CLAGNs with both the appearance/disappearance of broad emission line and the significant variation of $N_\mathrm{H}$ are much rarer \citep[][]{2022ApJ...927..227L}. The mechanism for the rapid variation of column density and the geometry of the torus for changing-obscuration AGNs are under debate. The possible association and discrepancy between the two types of changing-look phenomena are still unclear. The broadband X-ray spectral evolution of CLAGNs with both the variation of broad emission line and $N_\mathrm{H}$ might help us to explore these issues. The broadband X-ray spectra are crucial in determining the absorption and torus properties. In this work, we analyzed the Nuclear Spectroscopic Telescope Array \citep[\nustar\,;][]{2013ApJ...770..103H} spectra of five changing-look AGNs, which experienced both the rapid variation of the broad emission lines and the column density. In \autoref{sec:sample}, we describe the basic information of the sample. In \autoref{sec:model}, we present the data reduction and analysis. We simply discuss the possible explanation of the result in \autoref{sec:res} and \autoref{sec:dis} and summarize in \autoref{sec:sum}. Throughout this work, we adopt a flat $\Lambda-$CDM cosmological model with $H_0$=70 km s$^{-1}$ Mpc $^{-1}$, $\Omega_{m}$=0.27, and $\Omega_{\Lambda}$=$0.73$.

\section{Sample} \label{sec:sample}
The sample we collect meets the following criteria: (1) has experienced the appearance/disappearance of broad Balmer emission lines (2) extreme significant variation of the hydrogen column density in the obscured state (i.e., log$ \Delta (N_\mathrm{H}) > 1$ and log$ (N_\mathrm{H}/\,\mathrm{cm^{-2}}) > 22$) (3) at least one \nustar\, observation. Thus, ESO 362-G18, NGC 1365, NGC 4151, NGC 5548, and NGC 7582 are selected in our sample \citep[e.g.,][]{2007MNRAS.377..607P,2009ApJ...695..781B,2014MNRAS.443.2862A,2014ApJ...795...87B,2018Galax...6...52A,2019sf2a.conf..509M,2022ApJ...927..227L,2023MNRAS.518.2938T}.  3C 390.3 as a radio-loud galaxy is ruled out to avoid the jet effects. All of the sample are nearby and extremely variable Seyfert galaxies, thus abundant multi-epoch observations have been carried out to capture the changing-look events. All of them have a moderate disc inclination angle, and a relatively small BH mass (i.e., log($M_\mathrm{BH}/M_{\odot}$) between 6.65 and 7.74). For further study of the mechanism of changing-state and changing-obscuration phenomena, we take these five sources as a sample. The detailed information of the sample (i.e., \citealp[ESO 362-G18, ][]{2014MNRAS.443.2862A,2018Galax...6...52A}; \citealp[NGC 1365, ][]{2014ApJ...795...87B,2023MNRAS.518.2938T}; \citealp[NGC 4151, ][]{2007MNRAS.377..607P,2019sf2a.conf..509M}; \citealp[NGC 5548, ][]{2019ApJ...882L..30D,2014Sci...345...64K}; \citealp[NGC 7582, ][]{2009ApJ...695..781B,2019sf2a.conf..509M} ) is listed in \autoref{tab:info5} and presented in the appendix. 


\begin{table*}
\caption{Basic information of five CLAGNs. The table lists source name, redshift, the corresponding luminosity distance, the Galactic absorption ($N_{\rm{HI,Gal}}$) from \citet{Kalberla2005} adopted for X-ray spectral fitting, BH mass, inclination angle, and references. \\  \label{tab:info5}}
\centering
 \begin{tabular}{lc cc cc cc c} \hline\hline
{Name} & {Ref} &  {Redshift}&   {Luminosity distance}& {$N_{\rm{HI,Gal}}$}& {log($M_\mathrm{BH}/M_{\odot}$)}  & {Ref} & {disc inclination angle} & {Ref} \\ 
{} &{} &{} &{Mpc}  &{[10$^{20}$ cm$^{-2}$]} &{} &{} &{[deg]} & {} \\ \hline
ESO 362-G18  &  (1,2) &  0.0124 &  53.8 &  1.8 &  7.65 &  (1) &  $\sim 53\pm 5$ &  (1) \\
NGC 1365 &  (3,4) &  0.0055 &  23.5 &  1.3 &  6.65 &  (10) &  $\sim 57^{+3}_{-2}$ &  (14) \\
NGC 4151  &  (5,6) &  0.0033 &  14.3 &  2.3 &  7.6 &  (11) &  $\sim 58^{+8}_{-10}$ &  (15) \\
NGC 5548 &  (7,8) &  0.0172 &  74.6 &  1.6 &  7.7 &  (12) &  $\sim 39^{+12}_{-11}$ &  (16) \\
NGC 7582  &  (9,6) &  0.0053 &  22.6 &  1.3 &  7.74 &  (13) &  $\sim $68 &  (17) \\ \hline
\end{tabular}

References. (1)~\citet{2014MNRAS.443.2862A} (2)~\citet{2018Galax...6...52A} (3)~\citet{2014ApJ...795...87B} (4)~\citet{2023MNRAS.518.2938T} (5)~\citet{2007MNRAS.377..607P} (6)~\citet{2019sf2a.conf..509M} (7)~\citet{2019ApJ...882L..30D} (8)~\citet{2014Sci...345...64K} (9)~\citet{2009ApJ...695..781B} (10)~\citet{2017MNRAS.468L..97O} (11)~\citet{2022ApJ...936...75L} (12)~\citet{2015PASP..127...67B} (13)~\citet{2006A&A...460..449W} (14)~\citet{2010MNRAS.408..601W} (15)~\citet{2022ApJ...934..168B} (16)~\citet{2014MNRAS.445.3073P} (17)~\citet{Paturel2003} 
\end{table*}

\begin{table*}
\centering
\caption{The \nustar\,observations of five CLAGNs studied in this work. The table lists source name, \nustar\, observational ID, date, and exposure time. Observations with at least an exposure time of 1 ks are adopted.  We also include (quasi)-simultaneous soft X-ray data from \xmm\, and \xrt\, for part of \nustar\,observations. \label{tab:infonustar}}
\begin{tabular}{lc cc ccc} \hline \hline
{Name} & {ObsID}&  {MJD} & {Date} & {Exposure (ks)} \\ \hline
%
\multirow{2}*{ESO 362-G18}  & 60201046002 & 57655.8 & 2016-9-24 & 101.91 \\ 
                 & 790810101 (\xmm) &  &  2016-9-24 & \\ \hline
                 
 & 60002046002 & 56133.9 & 2012-7-25 & 36.26  \\
   & 692840201 (\xmm) &  & 2012-7-25 &   \\ 

\multirow{8}*{NGC 1365}  & 60002046003 & 56134.7 & 2012-7-26 & 40.59 \\
 & 35458003(\xrt) &  & 2012-7-26 &  \\

 & 60702058002 & 59320.9 & 2021-4-16 & 56.30 \\
     & 96123002(\xrt) &  & 2021-4-19 &  \\

 & 60702058004 & 59326.8 & 2021-4-22 & 38.05 \\
           & 89213003(\xrt)  &  & 2021-4-22 &  \\

 & 60702058006 & 59333.3 & 2021-4-29 & 11.85 \\
 & 60702058008 & 59336.0 & 2021-5-2 & 44.66 \\
 & 60702058010 & 59340.5 & 2021-5-6 & 41.48 \\
 & 60702058012 & 59342.7 & 2021-5-8 & 50.70 \\ \hline

  & 60001111002 & 56243.3 & 2012-11-12 & 21.86  \\
        & 80073001(\xrt) &  & 2012-11-12 &   \\
  & 60001111003 & 56243.8 & 2012-11-12 & 57.03  \\
\multirow{9}*{NGC 4151}  & 60001111005 & 56245.3 & 2012-11-14 & 61.53  \\
 & 60502017002 & 58688.3 & 2019-7-24 & 31.74 \\
 & 60502017004 & 58799.4 & 2019-11-12 & 43.74  \\
           & 88889001(\xrt) &  & 2019-11-12 &   \\
 & 60502017006 & 58841.5 & 2019-12-24 & 32.40  \\
  & 60502017008 & 58858.3 & 2020-1-10 & 30.59   \\
                        & 88889004(\xrt) &  & 2020-1-10 &   \\
  & 60502017010 & 58867.5 & 2020-1-19 & 29.72  \\
  & 60502017012 & 58871.3 & 2020-1-23 & 28.86  \\ 
            & 88889005(\xrt) &  & 2020-1-23 &   \\   \hline
& 60002044002 & 56484.4 & 2013-7-11 & 24.10 \\
     & 91744028(\xrt) &  & 2013-7-11 &   \\
\multirow{9}*{NGC 5548} & 60002044003 & 56485.0 & 2013-7-12 & 27.27 \\
      & 91711077(\xrt) &  & 2013-7-12 &   \\
 & 60002044005 & 56496.6 & 2013-7-23 & 49.52 \\
      & 91744040(\xrt) &  & 2013-7-23 &   \\
 & 60002044006 & 56545.9 & 2013-9-10 & 51.46 \\
      & 91711139(\xrt) &  & 2013-9-10 &   \\
 & 60002044008 & 56646.4 & 2013-12-20 & 50.10 \\
      & 80131001(\xrt) &  & 2013-12-21 &   \\
 & 90701601002 & 59240.3 & 2021-1-26 & 38.72 \\ 
      & 95671036(\xrt) &  & 2021-1-23 &   \\ \hline
      
  & 60061318002 & 56170.7 & 2012-8-31 & 16.46 \\
  & 32534001(\xrt) &  & 2012-9-1 &   \\
{NGC 7582}  & 60061318004 & 56184.7 & 2012-9-14 & 14.56 \\
 & 60201003002 & 57506.2 & 2016-4-28 & 48.49 \\
   & 782720301 (\xmm) &  & 2016-4-28 &   \\

\hline
\end{tabular}
\end{table*}

\section{Data reduction and analysis} \label{sec:model}
The \nustar\, FPMA/FPMB data were retrieved from \nustar\, Archive\footnote{\url{https://heasarc.gsfc.nasa.gov/docs/nustar/archive/nustar_archive.html}}  and reduced through the \texttt{nupipeline} task of the {\sc nustardas v2.1.2} package with the CALDB v20220316. The source and background spectra were extracted from a 90$''$ radius circle at the center of the source and a blank region, respectively. Each spectrum is grouped by a minimum of 30 counts per bin through the {\texttt grppha} command. The observational information of \nustar\, data is listed in \autoref{tab:infonustar}. We also reduced the soft X-ray data from \xmm\,/PN and \xrt\, for part of \nustar\, observations (see \autoref{tab:infonustar}).  The \xmm\,/PN data were reduced through Science Analysis Software (SAS, version 19.1.0). The source and background spectra were extracted through {\texttt epproc} from a circular region with a 40 arcsec radius that includes and excludes the source, respectively. The \xrt\, data in the photon-counting mode were reduced by {\sc xrtpipeline}. We extracted the spectra following the methods in \citet{2021MNRAS.506.4188L}.

There is only one \nustar\, observation for ESO 362-G18, which had low column density based on the Warm Corona and Relativistic Reflection models when combing the \xmm\, data \citep[e.g.,][]{2022RAA....22c5002Z}. For NGC 1365, the three \nustar\, spectra (ObsID 60002046005, 60002046007, and 60002046009) showed strong absorption iron lines around 6.7 keV and 7.0 keV during 2012 and 2013 \citep{2021RAA....21..199L}. For these spectra, we adopted the results from \citet{2021RAA....21..199L}. For NGC7582, \citet{2023MNRAS.522.1169L} performed a detailed time-resolved spectral analysis for the \nustar\, data in 2016 (ObsID 60201003002) based on a phenomenological model (xillver). The \nustar\, spectra are fitted in the 3--60 keV energy range with {\sc xspec v12.12.1} \citep{1996ASPC..101...17A} with the $\chi^{2}$ statistic. First, we fix the inclination angle to the values reported in \autoref{tab:info5}.

\subsection{pexrav}
We start the spectral fitting with a simple absorbed powerlaw model with exponential cut-off {\sc const * phabs1 * (zphabs2*cabs*zcutoffpl1}), where the {\sc const1} component represents the cross-calibration constants between different instruments, {\sc phabs1} represents the Galactic absorption, the additional {\sc zphabs2} and {\sc cabs} represent the line-of-sight absorption to account for the obscuration of the AGN. 
This simple model provides acceptable fits for ESO 362-G18 and the column density estimated is the same as that from the data constrained by only the \nustar\, data. The fitting is not quite acceptable for the other sources, so we add the reprocessed emission component modeled with \texttt{pexrav} \citep{1995MNRAS.273..837M} which represents the reprocessed X-ray emission from a cold, neutral, and semi-infinite slab. For ESO 362-G18, the added reflection component makes little difference to the fitting results. The iron abundance is set to one. We add a Gaussian line representing the Fe K$\alpha$ line. When the line width is not constrained, we fix it to 50 eV \citep[e.g.,][]{2022MNRAS.512.5942J}. When the cut-off energy of the primary component in \texttt{pexrav} is not well constrained, we fix it to 200 keV \citep[e.g.,][]{2022ApJ...930...46L} or link to the value of the observation with the longest exposure. For sources with high column densities, a scattered component is also usually observed. We then include one cutoff powerlaw component ({\sc const2*ZCUTOFFPL2}) representing the scattered component, where {\sc const2} represents the scattering fraction ($f_{\mathrm{s}}$). The X-ray photon index, cut-off energy, and normalization of the scattered component are linked to the primary continuum. The final model configuration used in {\sc xspec} is as follows, 
\begin{equation}\label{eq:slab}
\begin{aligned}
const1* phabs1 * (zphabs2*cabs*zcutoffpl1 \\ + zgauss + pexrav + const2*zcutoffpl2).
\end{aligned}
\end{equation}

The simple \texttt{pexrav} model provides good descriptions of the \nustar\, data. The parameters of the fitting results are listed in \autoref{tab:slab}. The intrinsic X-ray flux is calculated by {\it cflux} component. To study the spectral evolution and the variability of $N_\mathrm{H}$, the parameters we are mainly concerned about are the X-ray photon index ($\Gamma$), the column density of line-of-sight ($N_\mathrm{H, los}$), and the X-ray flux at $2-10$ keV. The intrinsic Eddington-scaled X-ray luminosity is estimated through error propagation based on the BH mass and the distance in \autoref{tab:info5}. The uncertainties of all spectral parameters are reported at the 90 percent confidence level if not otherwise specified. We also include the (quasi)-simultaneous \xmm\, or \xrt\, data for part of \nustar\, observations (see \autoref{tab:infonustar}) to better constrain the column density. 

The extreme variability of column density might be explained by the clumpy torus or the motion of obscuration material. We further consider two popular and physically motivated clumpy torus models (XCLUMPY and UXCLUMPY) to constrain the column density for the five CLAGNs. The best-fitted unfolded spectra with the \texttt{pexrav} model and two clumpy torus models are available in the appendix files.

\subsection{XCLUMPY}\label{sec:xclumpy}
As the changing look might be caused by the clumpy torus, we then test the widely used clumpy tours model \xclumpy\footnote{\url{https://github.com/AtsushiTanimoto/XClumpy}}~ \citep{2019ApJ...877...95T}. The \xclumpy\, model assumes a powerlaw distribution of clumps of the torus in the radial direction and a normal distribution in the elevation direction. The \xclumpy\, model has two components, {\sc xclumpy\_R} and {\sc xclumpy\_L}, which represent the reprocessed and line emission, respectively. The model configuration used in {\sc xspec} is as follows, 
\begin{equation}\label{eq:xclumpy}
\begin{aligned}
const1*phabs1*(zphabs2*cabs*zcutoffpl1\\ + xclumpy\_R   + const2*xclumpy\_L\\ + const3*zcutoffpl2).
\end{aligned}
\end{equation}
The equatorial column density, the inclination angle, and the torus angular width ($\sigma_{\rm tor}$) are free parameters, and the line-of-sight column density is linked with the above parameters in \xclumpy\, model as follows \citep{2019ApJ...877...95T},
\begin{equation}\label{eqn:xclumpy}
N_{\rm H}^{\rm los} = N_{\rm H}^{\rm Eq} [{\rm exp}~(\frac{(~i-\pi/2)^{2}}{\sigma_{\rm Tor}^{2}})].
\end{equation}
The photon index, cutoff energy, normalization of the primary component, the reprocessed and line component are linked together, and the other parameters of the reprocessed and line component are linked. The \xclumpy\, model considers a fixed cut-off energy at $E_{\rm cut}=370$ keV. The {\sc const2} and {\sc const3}  represent the relative normalization ($A_\mathrm{L}$) of the line emission and the scattering fraction ($f_\mathrm{s}$). We also try to constrain the inclination angles from the line of sight by setting it as a free parameter and jointly fitting the spectra. The parameters of the fitting results are presented in \autoref{tab:xclumpy}.

\label{sec:uxclumpy}
\subsection{UXCLUMPY}
Another clumpy torus model  {\uxcl\footnote{\url{https://github.com/JohannesBuchner/xars/blob/master/doc/uxclumpy.rst}}} \citep{2019A&A...629A..16B} is also tested. The \uxcl\, model assumes an axisymmetric and clumpy distribution of obscuring material. The number N of clouds between the observer and the central AGN is 
\begin{equation}\label{eq:uxclumpyN}
\begin{aligned}
N=N_{0} \times {\rm exp}\{-(\frac{\beta}{{\rm TOR}\sigma})\}^2 ,
\end{aligned}
\end{equation}
where $N_0$ is the number of clouds on the equatorial plane that remains constant, $\beta$ is the inclination angle toward the torus pole, and TOR$\sigma$ is the obscuring material angular width, which represents the torus scale height. Besides, the \uxcl\, model includes an inner Compton-thick ring of clouds with covering factor CTK (CTKcov) as a free parameter, and the column densities of these clouds are log$N_{\rm H}$=$25\pm0.5$. The column density of line-of-sight could be variable while the obscurer geometry keeps self-consistent. The model configuration used in {\sc xspec} is as follows, 
\begin{equation}\label{eq:uxclumpy}
\begin{aligned}
const1*phabs*(uxclumpy\_cutoff.fits +\\ f_{\rm s}*uxclumpy\_cutoff\_omni.fits),
\end{aligned}
\end{equation}
where \uxcl\, models the transmitted and the reflected component (including the fluorescent lines), and {\tt uxclumpy\_omni} models the so-called warm mirror emission (scattered emission). The scatter fraction is limited between 1e-5 and 0.1. The parameters of the \uxcl\, and {\tt uxclumpy\_omni} were linked. We froze the cutoff energy at 370 keV to be consistent with \xclumpy\,. We set CTKcov = 0.4 at first and then let CTKcov as a free parameter and refit the spectra. Similarly,  We also thaw the parameter of the inclination angle and refit the spectra. The parameters of the fitting results are presented in \autoref{tab:uxclumpy}.

\section{Results} \label{sec:res}
\begin{figure}
\centering
\includegraphics[width=0.35\textwidth]{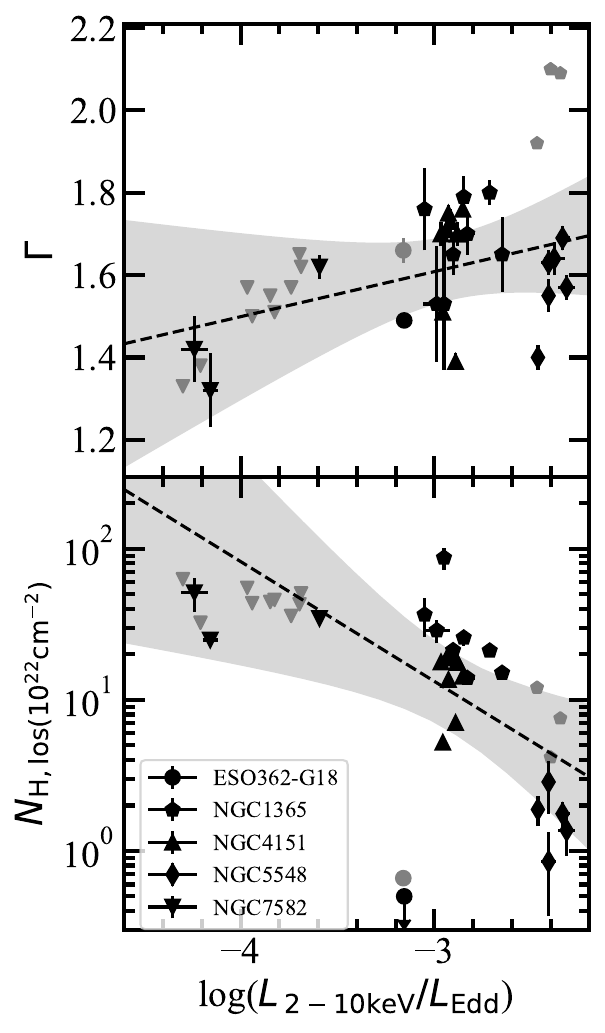}
\caption{Relations between $\Gamma$, log$N_{\rm H,los}$, and log$L_{\rm X}/L_{\rm Edd}$ based on \texttt{pexrav} model. The circle, pentagon, up-triangle, and down-triangle represent ESO 362-G18, NGC 1365, NGC4151, and NGC 7582. The dashed line represents the best-fit line with a 3 $\sigma$ confidence interval within the shaded region for all sources. Here, ESO 362-G18 is an obvious outlier in the log$N_{\rm H,los}$-log$L_{\rm X}L_{\rm Edd}$ correlation. The grey data points are from the literature for comparison (see \autoref{table:xray}). }
\label{pic:pexrav}
\end{figure}

The X-ray spectral photon indices are between $\sim1.4$ and $\sim2.0$, which are harder than other typical bright AGNs. The Eddington-scaled X-ray luminosity log$L_{\rm 2-10 keV}/L_{\rm Edd} $ range between $\sim -4.5$ and $\sim-2.5$.  Adopting the bolometric correction factors from \citet{2019MNRAS.488.5185N}, the Eddington ratio ($\lambda$ = log$L_{\rm bol}/L_{\rm Edd}$) range between $\sim -3.5$ and $\sim-1.0$. 

\subsection{Correlations}
To study the possible mechanism of the variability of the column density (i.e., obscuration effects or intrinsic accretion variation), we present the correlation between $\Gamma$, log$N_\mathrm{H,los}$, and log$L_{\rm X}/L_{\rm Edd}$ based on the \texttt{pexrav} model in \autoref{pic:pexrav}. We use the \texttt{linregress}\footnote{\textcolor{blue}{\url{https://docs.scipy.org/doc/scipy/reference/generated/scipy.stats.linregress.html}}} in the \texttt{scipy.stats} Python module to perform linear regression. The slope and intercept of the linear function, Pearson's correlation coefficient ($R-$value), and $p-$ values are reported in Table~\ref{tab:cor}. 

There is only one \nustar\, observation of ESO 362-G18 with low column density, which significantly deviates from the best-fit line for the log$N_\mathrm{H,los}$-log$L_{\rm X}/L_{\rm Edd}$ correlation. Then we exclude ESO 362-G18 during the linear fitting. We find a significant negative log$N_\mathrm{H,los}$-log$L_{\rm X}/L_{\rm Edd}$ correlation (slope $= -0.75 \pm 0.16 $), where the $R-$value is $-0.67 \pm 0.07 $ ($p = 6.7\times 10^{-5}$). For NGC 1365, we combine three \nustar\, observations, which show strong absorption iron lines \citep{2021RAA....21..199L}. For individual sources, the log$N_\mathrm{H,los}$ is also roughly anti-correlated with log$L_{\rm X}/L_{\rm Edd}$. For NGC 1365, there is a steeper correlation (slope $= -1.23$) with $R= -0.85 \pm 0.06 $ ($p = 9.3\times 10^{-4}$).  For NGC 7582, there are only three \nustar\, observations. We adopt the results from \citet{2023MNRAS.522.1169L}, which performs a detailed time-resolved spectral analysis of the \nustar\, observations in 2016 based on a phenomenological reflection model (see \autoref{table:xray}). NGC 7582 showed less variability of column density with a slope of -0.07 $\pm$ 0.15.

There is a positive correlation between $\Gamma$ and log$L_{\rm X}/L_{\rm Edd}$ (slope = $0.11 \pm 0.05$) for the whole sample, where the $R-$value is $0.39$ ($p $=$ 4.7\times 10^{-2}$). The positive correlation between $\Gamma$ and log$L_{\rm X}/L_{\rm Edd}$ also holds for NGC 1365, NGC 5548, and NGC 7582 with slopes of $0.68\pm 0.14$,  $1.29\pm 0.96$, and $0.46\pm 0.06$, respectively. For NGC 4151, the significance ($R$=$ 0.11\pm 0.21 $) is not so strong due to its less variability in X-ray luminosity.

\begin{figure*}
\centering
\begin{tikzpicture}
    \matrix[matrix of nodes]{
    \includegraphics[width=0.33\textwidth]{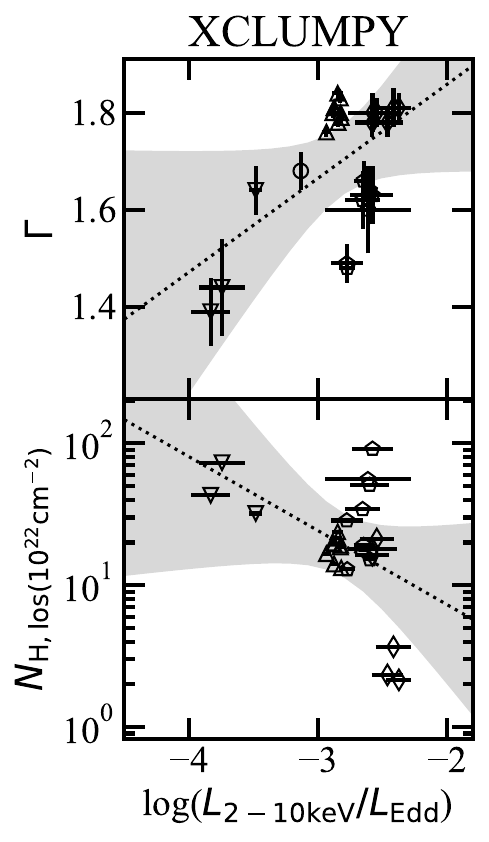} &
    \includegraphics[width=0.33\textwidth]{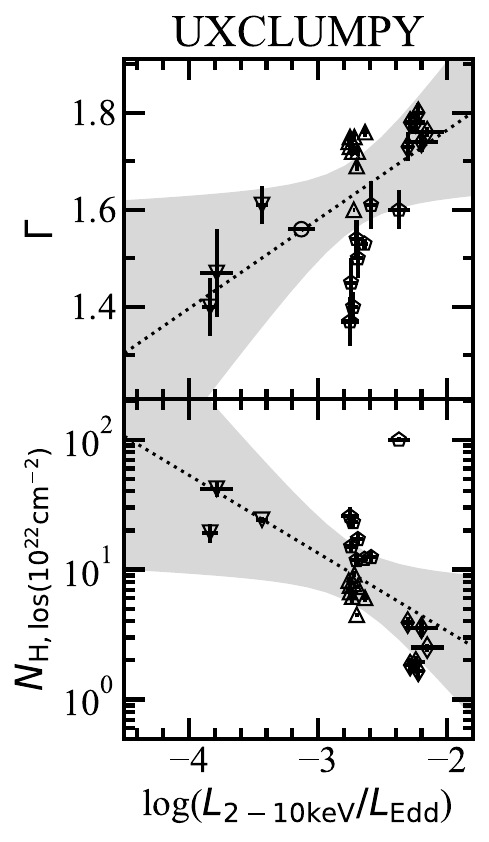} \\ 
    };
\end{tikzpicture}
\caption{Relations between $\Gamma$, log$N_{\rm H,los}$, and log$L_{\rm X}/L_{\rm Edd}$ for two clumpy torus models. }
\label{pic:torus_p4}
\end{figure*}

\begin{figure*}
\centering
\begin{tikzpicture}
    \matrix[matrix of nodes]{
    \includegraphics[width=0.33\textwidth]{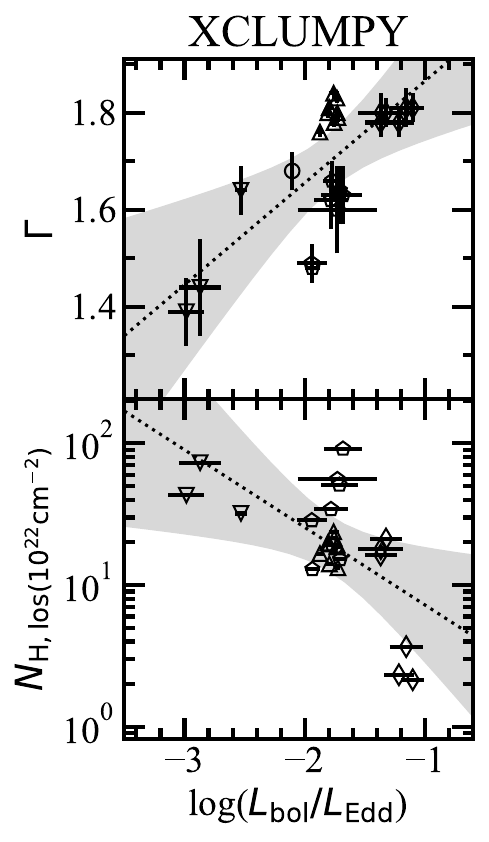} &
    \includegraphics[width=0.33\textwidth]{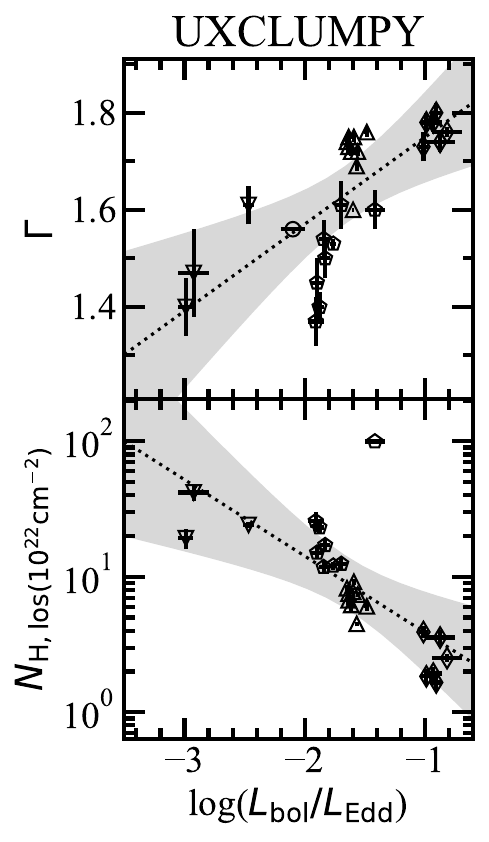} \\ 
    };
\end{tikzpicture}
\caption{Relations between $\Gamma$, log$N_{\rm H,los}$, and log$L_{\rm bol}/L_{\rm Edd}$ for two clumpy torus models. Bolometric correction factors are adopted from \citet{2019MNRAS.488.5185N}.}
\label{pic:torus_p5}
\end{figure*}

For further study and comparison, the correlations between $\Gamma$, log$N_\mathrm{H,los}$, and log$L_{\rm X}/L_{\rm Edd}$ based on two clumpy torus models are presented in \autoref{pic:torus_p4}. The positive correlation between $\Gamma$ and log$L_{\rm X}/L_{\rm Edd}$ for the five sources and the negative correlation between log$N_\mathrm{H,los}$ and log$L_{\rm X}/L_{\rm Edd}$ excluding ESO 362-G18 are further confirmed. 
The slopes of the $\Gamma$-log$L_{\rm X}/L_{\rm Edd}$ correlation and Pearson's correlation coefficients ($R-$value), and $p-$ values are 0.23 $\pm$ 0.03,  0.84 $\pm$ 0.04, and 4.1e-8 for the XCLUMPY model and 0.21 $\pm$ 0.02, 0.89 $\pm$ 0.03, and 3.2e-10 for the UXCLUMPY model, respectively. The slopes of the log$N_\mathrm{H,los}$-log$L_{\rm X}/L_{\rm Edd}$ correlation and Pearson's correlation coefficients ($R-$value), and $p-$ values are -0.56 $\pm$ 0.11,  -0.71 $\pm$ 0.07, and 5.2e-05 for the XCLUMPY model and -0.61 $\pm$ 0.08, -0.85 $\pm$ 0.04, and 3.9e-8 for the UXCLUMPY model, respectively. We also present the correlations between $\Gamma$, log$N_\mathrm{H,los}$, and log$L_{\rm bol}/L_{\rm Edd}$ for comparison in \autoref{pic:torus_p5}.

\begin{figure}
\centering
\includegraphics[width=0.4\textwidth]{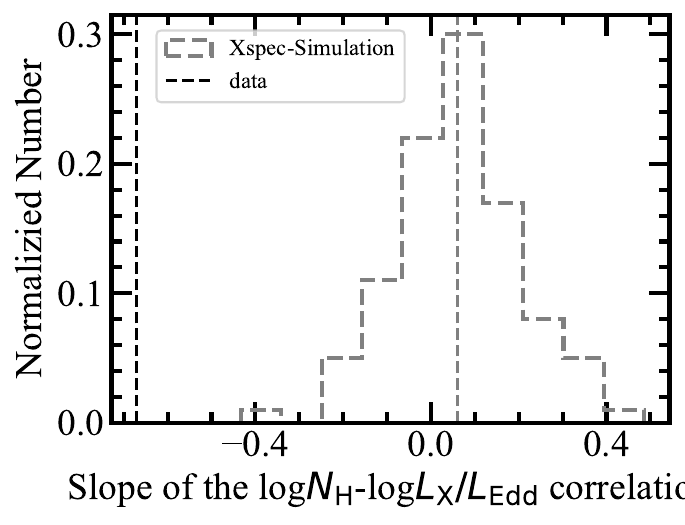} \\
\includegraphics[width=0.4\textwidth]{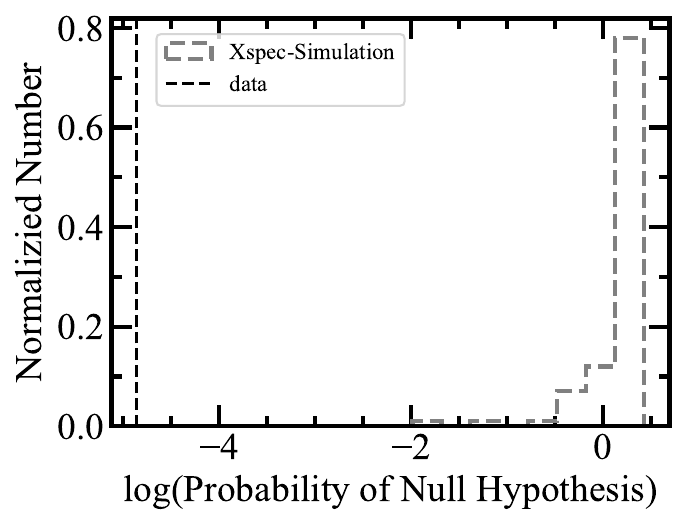}
\caption{Upper panel: the slope distribution of the correlation between log$N_{\rm H,los}$ and log$L_{\rm X}/L_{\rm Edd}$. Lower panel: the probability of
the null hypothesis. }
\label{pic:slope}
\end{figure}

\begin{table*}
\centering
\label{tab:cor}
\caption[]{Correlations for $\Gamma$-log$L_{\rm X}/L_{\rm Edd}$ and log$N_\mathrm{H,los}$-log$L_{\rm X}/L_{\rm Edd}$ based on the \texttt{pexrav} model. }

\begin{tabular}{lccc cccc} \hline\hline
{Sample} & {slope} & {intercept}  & {$R-$value}   & {$P-$value} \\ \hline
            &                  \multicolumn{3}{c} {$\Gamma$-log$L_{\rm X}/L_{\rm Edd}$   }                        &              & \\

All sample & 0.11 $\pm$ 0.05 & 1.94 $\pm$ 0.15 & 0.39 $\pm$ 0.11 & 4.7e-02 & \\   
ESO 362-G18 excluded  & 0.18 $\pm$ 0.06 & 2.19 $\pm$ 0.19 & 0.48 $\pm$ 0.10 & 8.5e-03 & \\
NGC 1365  & 0.68 $\pm$ 0.14 & 3.65 $\pm$ 0.38 & 0.85 $\pm$ 0.05 & 8.0e-04 & \\
NGC 4151 & 0.35 $\pm$ 1.24 & 2.68 $\pm$ 3.63 & 0.11 $\pm$ 0.22 & 7.9e-01 & \\
NGC 5548 & 1.29 $\pm$ 0.69 & 4.65 $\pm$ 1.65 & 0.68 $\pm$ 0.15 & 1.4e-01 & \\
NGC 7582 & 0.46 $\pm$ 0.06 & 3.31 $\pm$ 0.24 & 0.94 $\pm$ 0.03 & 1.5e-04 & \\
            &                                   \multicolumn{3}{c} { log$N_\mathrm{H,los}$-log$L_{\rm X}/L_{\rm Edd}$ }                         &              & \\         
ESO 362-G18 excluded & -0.75 $\pm$ 0.16 & -1.11 $\pm$ 0.46 & -0.67 $\pm$ 0.07 & 6.7e-05 & \\
NGC 1365  & -1.23 $\pm$ 0.25 & -2.11 $\pm$ 0.70 & -0.85 $\pm$ 0.06 & 9.3e-04 & \\
NGC 4151 & -0.47 $\pm$ 0.61 & -0.11 $\pm$ 1.76 & -0.28 $\pm$ 0.21 & 4.7e-01 & \\
NGC 5548 & -0.63 $\pm$ 1.86 & -1.34 $\pm$ 4.44 & -0.17 $\pm$ 0.27 & 7.5e-01 & \\
NGC 7582  & -0.07 $\pm$ 0.15 & 1.37 $\pm$ 0.59 & -0.18 $\pm$ 0.22 & 6.4e-01 & \\

\hline
\end{tabular}
\end{table*}

\subsection{Parameter degeneracy test}
To test the variability of log$N_\mathrm{H}$ dependent on log$L_{\rm X}/L_{\rm Edd}$ from the \texttt{pexrav} model is real rather than the degeneracy with parameter $\Gamma$,  we simulated 200 X-ray spectra (100 for FPMA and 100 for FPMB, respectively) for each observation of the obscured CLAGNs in \autoref{tab:infonustar} to test parameter degeneracies. The X-ray spectra are simulated with the {\sc fakeit} command using the same reflection model as an input. The input value of the column density from the line of sight is randomly selected from a uniform distribution of values between $10^{22} {\rm cm}^{-2}$  and $10^{24} {\rm cm}^{-2}$. The exposure time is set the same as the original observations. The input value of normalization is selected randomly from the distribution of those values from the previous fitting results.  The values for scattering fraction, $\Gamma$, high energy cutoff, and reflection factor are randomly selected from distributions following that in \citet[][]{2021MNRAS.504..428G}. For the remaining parameters such as redshift, iron abundance, and inclination, we fix them as before in the simulation and fitting. 

We present the slope distribution of the correlation between log$N_{\rm H,los}$ and log$L_{\rm X}/L_{\rm Edd}$  based on the fitting of the simulated spectra, see \autoref{pic:slope}. We adopt the fitting results with reduced $\chi^{2}<3 $. The negative slope for log$N_\mathrm{H}$-log$L_{\rm X}/L_{\rm Edd}$ correlation based on spectra fitting significantly deviates from that from simulations. The probability of the Null hypothesis is 1.4 $\times 10^{-5}$, which further confirms the validity of the reported negative correlation between log$N_{\rm H,los}$ and log$L_{\rm X}/L_{\rm Edd}$.

\section{Discussion} \label{sec:dis}
The changing-state and changing-obscuration phenomena could both be attributed to the change of accretion state or the obscuration effects \citep[e.g.,][]{2023NatAs...7.1282R}. However, the mechanism for the simultaneous variation of X-ray spectra and the column density is under debate. The smooth or clumpy nature of the torus for those changing-look AGNs is still unclear. The rapid variation of X-ray spectra and the column density from the line of sight within a short timescale can also be attributed to the variable obscuration or clumpy nature of absorbing gas for NGC 1365 \citep{2013MNRAS.429.2662B} and ESO 362-G18 \citep[][]{2014MNRAS.443.2862A}. One recent work also supports that a fully covering clumpy absorber model could better explain the long-term and short-term absorption variability in NGC 7582 compared to the phenomenological reflection model \citep[see][]{2023MNRAS.522.1169L}. For NGC 5548, the spectral changes could be explained by the emergence of disk wind \citep[][]{2019ApJ...882L..30D,2019ApJ...881..153K,2023NatAs...7.1282R}.

Some other scenarios have also been proposed that the variability of the line-of-sight column density is driven by the change of intrinsic luminosity or the ionization state, which also has a significant impact on the variation of broad emission lines in the BLR as seen in the optical CLAGNs \citep[e.g.,][]{2015OAP....28..175O,2022AN....34310080O}. One possible scenario is that the variation of intrinsic ionizing luminosity could cause the sublimation/condensation of dusty clouds near the central engine \citep[e.g.,][]{2021MNRAS.507..687J,2022AN....34310080O}. The sublimation timescale ($t_{sub} \sim 10^{3-5} {\rm s}$ ) is much shorter than the typical recovery timescale ($\sim$ several years). The typical optical changing-look AGN NGC 1566 experienced the variation of column density during the 2018 outburst  \citep[$N_\mathrm{H,los}  \sim 3.5 \times 10^{21} {\rm cm^{-2}} $ pre-outburst and recovered from $\sim 6$ $\times 10^{20} {\rm cm^{-2}} $ to 1.3 $\times 10^{21} {\rm cm^{-2}}$ after the outburst; see ][]{2021MNRAS.507..687J}. The unobscured nature ($N_\mathrm{H,los} < 10^{22} {\rm cm^{-2}} $ ) of NGC 1566 could be attributed to its face-on view with an inclination angle of $\sim 10-18 ^{\circ}$ \citep{2019MNRAS.483..558O}. The increase of ionizing luminosity could lead to the dramatic enhancement of broad line emission and the variability of absorption column density through the sublimation of dust \citep{2019MNRAS.483..558O}. In this scenario, \citet{2019MNRAS.483..558O} predicts that the $N_\mathrm{H,los}$ would recover to the level before the outburst in several years. 

Another scenario is that with the increase of the accretion rate and the ionizing luminosity, the increase of the ionization state of the obscuring gas could lead to a decrease in column density. This scenario is originally proposed for NGC 4151 \citep{1989MNRAS.236..153Y}. NGC 4151 experienced significant intrinsic changes in the X-ray luminosity and spectral shape \citep[e.g.,][]{1989MNRAS.236..153Y,2007MNRAS.377..607P,2017A&A...603A..50B}. The rapid variability of absorption is consistent with the obscuring material located at BLR \citep[see][]{2007MNRAS.377..607P}. Besides, the dust reverberation mapping result for NGC4151 is inconsistent with the prediction of the pure clumpy torus model \citep[][]{2021ApJ...912..126L}. But a larger amplitude of variations in the luminosity is required for NGC 1365 and other changing-obscuration AGNs than observations in the variable ionization state scenario \citep{2005ApJ...623L..93R,2023NatAs...7.1282R}.

There is a widely existing positive correlation between the X-ray photon index ($\Gamma$) and the Eddington-scaled X-ray $2-10$ keV luminosity (log$L_{\rm X}$/$L_{\rm Edd}$, hereafter) to trace the spectral evolution of the corona in bright AGNs \citep[e.g., ][]{2017MNRAS.470..800T}. A positive $\Gamma$ - log$L_{\rm X}$/$L_{\rm Edd}$ correlation (slope $\sim 0.19$) is also found for the five sources, which is consistent with the positive $\Gamma$-log$L_{\rm X}$/$L_{\rm Edd}$ correlation with different slopes varying from $0.1$ to $0.3$ for individual sources and different samples \citep[e.g.,][]{2017MNRAS.470..800T,2018MNRAS.480.1819R}. The positive $\Gamma$-log$L_{\rm X}/L_{\rm Edd}$ correlation can be attributed to the seed photons of Compton scattering coming from the thermal emission of SSD \citep[e.g., ][]{2013ApJ...764....2Q}. In the disc-corona system, the cooling of the electron temperature is more effective and the X-ray spectra become softer with the increase of the accretion rate. The result supports that the five CLAGNs with strong variations of column density are in the SSD (intrinsically type 1) accretion state \citep{2021MNRAS.506.4188L}.

Four of them (i.e., NGC 1365, NGC 4151, NGC 5548, and NGC 7582) produce strong variations of the intrinsic hard X-ray luminosities, which cannot be explained only by the variation of obscuration. The MIR emission is mainly produced by the hot dust heated by the UV radiation from the accretion disc \citep{2018ARA&A..56..625H}, which is less influenced by the obscuration effects. These sources also experienced significant MIR magnitude variabilities based on {\it WISE} data with $\Delta W$1 of 0.25, 1.13, 0.47, and 0.55 for NGC 1365, NGC 4151, NGC 5548, and NGC 7582, respectively \citep[][]{2022ApJ...927..227L}. The MIR magnitude variation of the three AGNs is much larger than normal AGNs \citep[$\Delta W$1 $\sim$ 0.1;][]{2022ApJ...927..227L}, which also suggests that the extreme variability of column density can not be simply caused by the obscuration effect.

We find that the line-of-sight column density ($N_\mathrm{H,los}$) declined with the increase of the Eddington-scaled X-ray luminosity ($L_{\rm X}/L_\mathrm{Edd}$) for the whole sample except for ESO 362-G18. This suggests that the variation of column density for three CLAGNs is directly regulated by the intrinsic ionizing luminosity (i.e. accretion rate). The variation of the column density might be driven by the variable winds/outflows forming from the disc at moderate inclination angles (e.g., $\sim 40^\circ-70^\circ$) from our line of sight \citep[e.g.,][]{2020MNRAS.492.5540M}. The disc winds with a certain open angle can push the obscuring material away along the direction of the line of sight and then decrease the column density \citep[e.g.,][]{2022A&A...662A..77M}. The strength of radiation pressure-driven wind positively correlates with the accretion rate. The negative log$N_{\rm H,los}$-log$L_{\rm X}/L_{\rm Edd}$ correlation supports the disc wind scenario as the explanation of the variability of line-of-sight column density in our sample with moderate inclination angles except for ESO 362-G18. In the scenario of radiation-pressure-driven blowout region \citep[e.g.,][]{2017Natur.549..488R}, AGNs with outflows have a higher Eddington ratio between $10^{-1.5}$ and $10^{0}$. In our sample, NGC 1365, NGC 4151, and NGC 7582 showed significant negative log$N_{\rm H,los}$-log$L_{\rm X}/L_{\rm Edd}$ correlation and had a value of Eddington ratio larger than $10^{-2}$. NGC 7582 in the low state had an Eddington ratio smaller than $10^{-2}$ and it had the lowest variability of column density (see \autoref{pic:torus_p5}), which is roughly consistent with the radiation-regulated unification of AGN \citep[see Figure 4 in][]{2017Natur.549..488R}. We notice that the slope of log$N_\mathrm{H,los}$-log$L_{\rm X}/L_\mathrm{Edd}$ correlation for NGC 1365 is steeper than that of the other sources, which implies the stronger disc wind in NGC 1365.

\begin{figure*}
\centering
\includegraphics[width=0.8\textwidth]{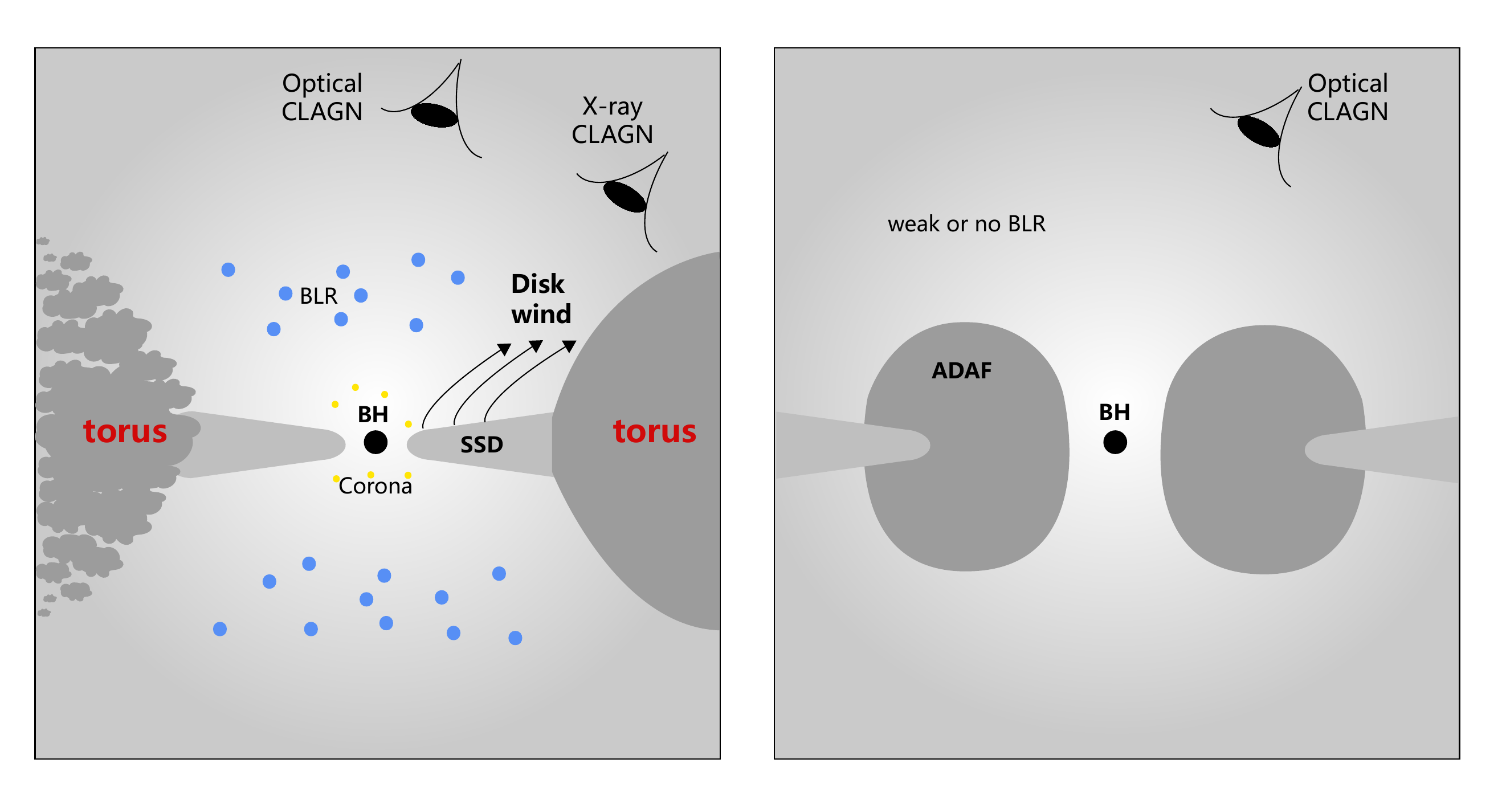}
\caption{A unified scenario for two kinds of changing-look phenomena. The dusty torus could consist of smooth or clumpy clouds. The optical CLAGNs are face-viewed with low-to-intermediate inclination angles and experience the transition between SSD and ADAF state. The variability of column density in X-ray CLAGNs should be regulated not only by the clumpy torus but also by the disk wind formed at intermediate inclination angles with the change of accretion rate in the SSD state.}  
\label{pic:clagn}
\end{figure*}

The hard X-ray observations support the unified scheme for the two kinds of changing-look phenomena (see \autoref{pic:clagn}), which could also naturally explain the observed multi-wavelength (optical/MIR) properties for changing-look AGNs. Similar to changing-state AGNs, changing-obscuration AGNs may be also triggered by the evolution of the accretion disc. Face-on sources with low inclination angles should show little variability of the low column density. For edge-on sources with high inclination angles, the BLRs are fully obscured by the torus. Only sources with moderate inclination angles are expected to show the two kinds of changing-look phenomena, which is consistent with our sample. The disk wind scenario should be only valid in the intrinsically type 1 state (see left part of \autoref{pic:clagn}). When the accretion rate decreases, the ioning luminosity and the disk wind would be weaker, the obscuration would increase, and the optical state would also transit into the 1.5 or even 1.9 state. In the type 2 state, the plasma is fully ionized by the hot electrons in the ADAF case \citep[e.g.,][]{2022iSci...25j3544L} and the wind may have little influence on the column density (see right part of \autoref{pic:clagn}).

In such a scenario, the variation of the column density should be correlated with the variation of intrinsic luminosity in the smooth torus. For smooth torus, the variations of $\Gamma$ and $N_\mathrm{H,los}$ are both dependent on the change of accretion rate, which is correlated with the strength of the wind. The sources would not experience the abrupt variation of $N_\mathrm{H,los}$ while the intrinsic spectra and luminosity do not change a lot. ESO 362-G18 is an outlier in the log$N_\mathrm{H,los}$-log$L_{\rm X}/L_\mathrm{Edd}$ correlation, which is consistent with that the variable absorber is attributed to a dusty and clumpy torus \citep[][]{2014MNRAS.443.2862A}. Thus, the log$N_{\rm H,los}$-log$L_{\rm X}/L_{\rm Edd}$ and $\Gamma$-log$L_{\rm X}/L_{\rm Edd}$ correlations could be used as a probe for testing the clumpy torus nature for AGNs with extreme variability of column density.

\section{Conclusions} \label{sec:sum}
In this work, we explore the X-ray spectral evolution and the variability of column density for five changing-look AGNs based on \nustar\, data. All five sources have experienced two types of changing-look phenomena and have a moderate inclination angle.  We summarize the main results as follows:

Based on a phenomenological reflection model \texttt{pexrav} and two clumpy torus models,  a significant negative log$N_{\rm H,los}$-log$L_{\rm X}/L_{\rm Edd}$ correlation is found for NGC 1365, NGC 4151, NGC 5548, and NGC 7582. This result supports that the variation of the hydrogen absorption column density is regulated by the variation of intrinsic ionizing luminosity rather than the simple obscuration effect. The variation of the column density might be triggered by the variable wind forming in the disc when observed at an appropriate inclination angle from the line of sight.  ESO 362-G18 is an obvious outlier in the log$N_{\rm H,los}$-log$L_{\rm X}/L_{\rm Edd}$ correlation, which could be explained by its clumpy torus structure. 

The positive $\Gamma$-log$L_{\rm X}/L_{\rm Edd}$ correlation supports their SSD accretion mode. For NGC 1365, the slope of the $\Gamma$-log$L_{\rm X}/L_{\rm Edd}$ correlation is significantly steeper than that of the whole sample. Similarly and interestingly, the slope of the negative log$N_{\rm H,los}$-log$L_{\rm X}/L_{\rm Edd}$ correlation is also significantly steeper for NGC 1365 compared to the whole sample, which implies the stronger disc wind formed in NGC 1365. 

The changing-obscuration phenomena are not only modulated by its clumpy torus nature but also mostly possibly triggered by the wind forming in the disc during the evolution of the accretion state with variable accretion rate.

\section*{Acknowledgements}
We are thankful for the support of the National Science Foundation of China (11721303, 11927804, and 12133001) and the National Key R\&D Program of China (2022YFF0503401). We acknowledge the science research grant from the China Manned Space Project with No. CMS-CSST-2021-A06. QW was supported in part by the Natural Science Foundation of China (grant U1931203); ZY was supported in part by the Natural Science Foundation of China (grants U1938114), the Youth Innovation Promotion Association of CAS (id 2020265) and funds for key programs of Shanghai astronomical observatory; and WY would like to acknowledge the support in part by the National Program on Key Research and Development Project (grant 2016YFA0400804) and the National Natural Science Foundation of China (grants 11333005 and U1838203). This research has made use of the NuSTAR Data Analysis Software (NuSTARDAS) package jointly developed by the ASI Space Science Data Center (SSDC, Italy) and the California Institute of Technology (Caltech, USA), and the data provided by the High Energy Astrophysics Science Archive Research Center (HEASARC). This research has made use of NASA's Astrophysics Data System (ADS) and NASA/IPAC  EXTRAGALACTIC  DATABASE (NED). 


\section*{Data Availability}
The data underlying this article are available through HEASARC Browse database.



\bibliographystyle{mnras}
\bibliography{ref} 

\begin{thebibliography}{}
\makeatletter
\relax
\def\mn@urlcharsother{\let\do\@makeother \do\$\do\&\do\#\do\^\do\_\do\%\do\~}
\def\mn@doi{\begingroup\mn@urlcharsother \@ifnextchar [ {\mn@doi@}
  {\mn@doi@[]}}
\def\mn@doi@[#1]#2{\def\@tempa{#1}\ifx\@tempa\@empty \href
  {http://dx.doi.org/#2} {doi:#2}\else \href {http://dx.doi.org/#2} {#1}\fi
  \endgroup}
\def\mn@eprint#1#2{\mn@eprint@#1:#2::\@nil}
\def\mn@eprint@arXiv#1{\href {http://arxiv.org/abs/#1} {{\tt arXiv:#1}}}
\def\mn@eprint@dblp#1{\href {http://dblp.uni-trier.de/rec/bibtex/#1.xml}
  {dblp:#1}}
\def\mn@eprint@#1:#2:#3:#4\@nil{\def\@tempa {#1}\def\@tempb {#2}\def\@tempc
  {#3}\ifx \@tempc \@empty \let \@tempc \@tempb \let \@tempb \@tempa \fi \ifx
  \@tempb \@empty \def\@tempb {arXiv}\fi \@ifundefined
  {mn@eprint@\@tempb}{\@tempb:\@tempc}{\expandafter \expandafter \csname
  mn@eprint@\@tempb\endcsname \expandafter{\@tempc}}}

\bibitem[\protect\citeauthoryear{{Ag{\'\i}s-Gonz{\'a}lez}
  et~al.,}{{Ag{\'\i}s-Gonz{\'a}lez} et~al.}{2014}]{2014MNRAS.443.2862A}
{Ag{\'\i}s-Gonz{\'a}lez} B.,  et~al., 2014, \mn@doi [\mnras]
  {10.1093/mnras/stu1358}, \href
  {https://ui.adsabs.harvard.edu/abs/2014MNRAS.443.2862A} {443, 2862}

\bibitem[\protect\citeauthoryear{{Ag{\'\i}s-Gonz{\'a}lez}, {Hutsem{\'e}kers}
  \& {Miniutti}}{{Ag{\'\i}s-Gonz{\'a}lez} et~al.}{2018}]{2018Galax...6...52A}
{Ag{\'\i}s-Gonz{\'a}lez} B.,  {Hutsem{\'e}kers} D.,   {Miniutti} G.,  2018,
  \mn@doi [Galaxies] {10.3390/galaxies6020052}, \href
  {https://ui.adsabs.harvard.edu/abs/2018Galax...6...52A} {6, 52}

\bibitem[\protect\citeauthoryear{{Andonie} et~al.,}{{Andonie}
  et~al.}{2022}]{2022A&A...664A..46A}
{Andonie} C.,  et~al., 2022, \mn@doi [\aap] {10.1051/0004-6361/202142473},
  \href {https://ui.adsabs.harvard.edu/abs/2022A&A...664A..46A} {664, A46}

\bibitem[\protect\citeauthoryear{{Antonucci}}{{Antonucci}}{1993}]{1993ARA&A..31..473A}
{Antonucci} R.,  1993, \mn@doi [\araa] {10.1146/annurev.aa.31.090193.002353},
  \href {https://ui.adsabs.harvard.edu/abs/1993ARA&A..31..473A} {31, 473}

\bibitem[\protect\citeauthoryear{{Arnaud}}{{Arnaud}}{1996}]{1996ASPC..101...17A}
{Arnaud} K.~A.,  1996, in {Jacoby} G.~H.,  {Barnes} J.,  eds,  Astronomical
  Society of the Pacific Conference Series Vol. 101, Astronomical Data Analysis
  Software and Systems V. p.~17

\bibitem[\protect\citeauthoryear{{Balokovi{\'c}} et~al.,}{{Balokovi{\'c}}
  et~al.}{2018}]{2018ApJ...854...42B}
{Balokovi{\'c}} M.,  et~al., 2018, \mn@doi [\apj] {10.3847/1538-4357/aaa7eb},
  \href {https://ui.adsabs.harvard.edu/abs/2018ApJ...854...42B} {854, 42}

\bibitem[\protect\citeauthoryear{{Bentz} \& {Katz}}{{Bentz} \&
  {Katz}}{2015}]{2015PASP..127...67B}
{Bentz} M.~C.,  {Katz} S.,  2015, \mn@doi [\pasp] {10.1086/679601}, \href
  {https://ui.adsabs.harvard.edu/abs/2015PASP..127...67B} {127, 67}

\bibitem[\protect\citeauthoryear{{Bentz}, {Williams}  \& {Treu}}{{Bentz}
  et~al.}{2022}]{2022ApJ...934..168B}
{Bentz} M.~C.,  {Williams} P.~R.,   {Treu} T.,  2022, \mn@doi [\apj]
  {10.3847/1538-4357/ac7c0a}, \href
  {https://ui.adsabs.harvard.edu/abs/2022ApJ...934..168B} {934, 168}

\bibitem[\protect\citeauthoryear{{Beuchert} et~al.,}{{Beuchert}
  et~al.}{2017}]{2017A&A...603A..50B}
{Beuchert} T.,  et~al., 2017, \mn@doi [\aap] {10.1051/0004-6361/201630293},
  \href {https://ui.adsabs.harvard.edu/abs/2017A&A...603A..50B} {603, A50}

\bibitem[\protect\citeauthoryear{{Bianchi}, {Piconcelli}, {Chiaberge},
  {Bail{\'o}n}, {Matt}  \& {Fiore}}{{Bianchi}
  et~al.}{2009}]{2009ApJ...695..781B}
{Bianchi} S.,  {Piconcelli} E.,  {Chiaberge} M.,  {Bail{\'o}n} E.~J.,  {Matt}
  G.,   {Fiore} F.,  2009, \mn@doi [\apj] {10.1088/0004-637X/695/1/781}, \href
  {https://ui.adsabs.harvard.edu/abs/2009ApJ...695..781B} {695, 781}

\bibitem[\protect\citeauthoryear{{Braito}, {Reeves}, {Gofford}, {Nardini},
  {Porquet}  \& {Risaliti}}{{Braito} et~al.}{2014}]{2014ApJ...795...87B}
{Braito} V.,  {Reeves} J.~N.,  {Gofford} J.,  {Nardini} E.,  {Porquet} D.,
  {Risaliti} G.,  2014, \mn@doi [\apj] {10.1088/0004-637X/795/1/87}, \href
  {https://ui.adsabs.harvard.edu/abs/2014ApJ...795...87B} {795, 87}

\bibitem[\protect\citeauthoryear{{Brenneman}, {Risaliti}, {Elvis}  \&
  {Nardini}}{{Brenneman} et~al.}{2013}]{2013MNRAS.429.2662B}
{Brenneman} L.~W.,  {Risaliti} G.,  {Elvis} M.,   {Nardini} E.,  2013, \mn@doi
  [\mnras] {10.1093/mnras/sts555}, \href
  {https://ui.adsabs.harvard.edu/abs/2013MNRAS.429.2662B} {429, 2662}

\bibitem[\protect\citeauthoryear{{Buchner}, {Brightman}, {Nandra}, {Nikutta}
  \& {Bauer}}{{Buchner} et~al.}{2019}]{2019A&A...629A..16B}
{Buchner} J.,  {Brightman} M.,  {Nandra} K.,  {Nikutta} R.,   {Bauer} F.~E.,
  2019, \mn@doi [\aap] {10.1051/0004-6361/201834771}, \href
  {https://ui.adsabs.harvard.edu/abs/2019A&A...629A..16B} {629, A16}

\bibitem[\protect\citeauthoryear{{Chen} et~al.,}{{Chen}
  et~al.}{2023}]{2023MNRAS.520.1807C}
{Chen} Y.-J.,  et~al., 2023, \mn@doi [\mnras] {10.1093/mnras/stad051}, \href
  {https://ui.adsabs.harvard.edu/abs/2023MNRAS.520.1807C} {520, 1807}

\bibitem[\protect\citeauthoryear{{Dehghanian} et~al.,}{{Dehghanian}
  et~al.}{2019}]{2019ApJ...882L..30D}
{Dehghanian} M.,  et~al., 2019, \mn@doi [\apjl] {10.3847/2041-8213/ab3d41},
  \href {https://ui.adsabs.harvard.edu/abs/2019ApJ...882L..30D} {882, L30}

\bibitem[\protect\citeauthoryear{{Denney} et~al.,}{{Denney}
  et~al.}{2014}]{2014ApJ...796..134D}
{Denney} K.~D.,  et~al., 2014, \mn@doi [\apj] {10.1088/0004-637X/796/2/134},
  \href {https://ui.adsabs.harvard.edu/abs/2014ApJ...796..134D} {796, 134}

\bibitem[\protect\citeauthoryear{{Di Gesu} et~al.,}{{Di Gesu}
  et~al.}{2015}]{2015A&A...579A..42D}
{Di Gesu} L.,  et~al., 2015, \mn@doi [\aap] {10.1051/0004-6361/201525934},
  \href {https://ui.adsabs.harvard.edu/abs/2015A&A...579A..42D} {579, A42}

\bibitem[\protect\citeauthoryear{{Elitzur}, {Ho}  \& {Trump}}{{Elitzur}
  et~al.}{2014}]{2014MNRAS.438.3340E}
{Elitzur} M.,  {Ho} L.~C.,   {Trump} J.~R.,  2014, \mn@doi [\mnras]
  {10.1093/mnras/stt2445}, \href
  {https://ui.adsabs.harvard.edu/abs/2014MNRAS.438.3340E} {438, 3340}

\bibitem[\protect\citeauthoryear{{Fraquelli}, {Storchi-Bergmann}  \&
  {Binette}}{{Fraquelli} et~al.}{2000}]{2000ApJ...532..867F}
{Fraquelli} H.~A.,  {Storchi-Bergmann} T.,   {Binette} L.,  2000, \mn@doi
  [\apj] {10.1086/308621}, \href
  {https://ui.adsabs.harvard.edu/abs/2000ApJ...532..867F} {532, 867}

\bibitem[\protect\citeauthoryear{{Gandhi}, {H{\"o}nig}  \&
  {Kishimoto}}{{Gandhi} et~al.}{2015}]{2015ApJ...812..113G}
{Gandhi} P.,  {H{\"o}nig} S.~F.,   {Kishimoto} M.,  2015, \mn@doi [\apj]
  {10.1088/0004-637X/812/2/113}, \href
  {https://ui.adsabs.harvard.edu/abs/2015ApJ...812..113G} {812, 113}

\bibitem[\protect\citeauthoryear{{Grier} et~al.,}{{Grier}
  et~al.}{2013}]{2013ApJ...773...90G}
{Grier} C.~J.,  et~al., 2013, \mn@doi [\apj] {10.1088/0004-637X/773/2/90},
  \href {https://ui.adsabs.harvard.edu/abs/2013ApJ...773...90G} {773, 90}

\bibitem[\protect\citeauthoryear{{Guolo}, {Ruschel-Dutra}, {Grupe}, {Peterson},
  {Storchi-Bergmann}, {Schimoia}, {Nemmen}  \& {Robinson}}{{Guolo}
  et~al.}{2021}]{2021MNRAS.508..144G}
{Guolo} M.,  {Ruschel-Dutra} D.,  {Grupe} D.,  {Peterson} B.~M.,
  {Storchi-Bergmann} T.,  {Schimoia} J.,  {Nemmen} R.,   {Robinson} A.,  2021,
  \mn@doi [\mnras] {10.1093/mnras/stab2550}, \href
  {https://ui.adsabs.harvard.edu/abs/2021MNRAS.508..144G} {508, 144}

\bibitem[\protect\citeauthoryear{{Gupta} et~al.,}{{Gupta}
  et~al.}{2021}]{2021MNRAS.504..428G}
{Gupta} K.~K.,  et~al., 2021, \mn@doi [\mnras] {10.1093/mnras/stab839}, \href
  {https://ui.adsabs.harvard.edu/abs/2021MNRAS.504..428G} {504, 428}

\bibitem[\protect\citeauthoryear{{Harrison} et~al.,}{{Harrison}
  et~al.}{2013}]{2013ApJ...770..103H}
{Harrison} F.~A.,  et~al., 2013, \mn@doi [\apj] {10.1088/0004-637X/770/2/103},
  \href {https://ui.adsabs.harvard.edu/abs/2013ApJ...770..103H} {770, 103}

\bibitem[\protect\citeauthoryear{{Hickox} \& {Alexander}}{{Hickox} \&
  {Alexander}}{2018}]{2018ARA&A..56..625H}
{Hickox} R.~C.,  {Alexander} D.~M.,  2018, \mn@doi [\araa]
  {10.1146/annurev-astro-081817-051803}, \href
  {https://ui.adsabs.harvard.edu/abs/2018ARA&A..56..625H} {56, 625}

\bibitem[\protect\citeauthoryear{{Ho}}{{Ho}}{2008}]{2008ARA&A..46..475H}
{Ho} L.~C.,  2008, \mn@doi [\araa] {10.1146/annurev.astro.45.051806.110546},
  \href {https://ui.adsabs.harvard.edu/abs/2008ARA&A..46..475H} {46, 475}

\bibitem[\protect\citeauthoryear{{Jana}, {Kumari}, {Nandi}, {Naik},
  {Chatterjee}, {Jaisawal}, {Hayasaki}  \& {Ricci}}{{Jana}
  et~al.}{2021}]{2021MNRAS.507..687J}
{Jana} A.,  {Kumari} N.,  {Nandi} P.,  {Naik} S.,  {Chatterjee} A.,  {Jaisawal}
  G.~K.,  {Hayasaki} K.,   {Ricci} C.,  2021, \mn@doi [\mnras]
  {10.1093/mnras/stab2155}, \href
  {https://ui.adsabs.harvard.edu/abs/2021MNRAS.507..687J} {507, 687}

\bibitem[\protect\citeauthoryear{{Jana} et~al.,}{{Jana}
  et~al.}{2022}]{2022MNRAS.512.5942J}
{Jana} A.,  et~al., 2022, \mn@doi [\mnras] {10.1093/mnras/stac799}, \href
  {https://ui.adsabs.harvard.edu/abs/2022MNRAS.512.5942J} {512, 5942}

\bibitem[\protect\citeauthoryear{{Jana} et~al.,}{{Jana}
  et~al.}{2024}]{2024arXiv241108676J}
{Jana} A.,  et~al., 2024, \mn@doi [arXiv e-prints] {10.48550/arXiv.2411.08676},
  \href {https://ui.adsabs.harvard.edu/abs/2024arXiv241108676J} {p.
  arXiv:2411.08676}

\bibitem[\protect\citeauthoryear{{Jin}, {Wu}  \& {Feng}}{{Jin}
  et~al.}{2022}]{2022ApJ...926..184J}
{Jin} J.-J.,  {Wu} X.-B.,   {Feng} X.-T.,  2022, \mn@doi [\apj]
  {10.3847/1538-4357/ac410c}, \href
  {https://ui.adsabs.harvard.edu/abs/2022ApJ...926..184J} {926, 184}

\bibitem[\protect\citeauthoryear{{Kaastra} et~al.,}{{Kaastra}
  et~al.}{2014}]{2014Sci...345...64K}
{Kaastra} J.~S.,  et~al., 2014, \mn@doi [Science] {10.1126/science.1253787},
  \href {https://ui.adsabs.harvard.edu/abs/2014Sci...345...64K} {345, 64}

\bibitem[\protect\citeauthoryear{{Kalberla}, {Burton}, {Hartmann}, {Arnal},
  {Bajaja}, {Morras}  \& {P{\"o}ppel}}{{Kalberla} et~al.}{2005}]{Kalberla2005}
{Kalberla} P.~M.~W.,  {Burton} W.~B.,  {Hartmann} D.,  {Arnal} E.~M.,  {Bajaja}
  E.,  {Morras} R.,   {P{\"o}ppel} W.~G.~L.,  2005, \mn@doi [\aap]
  {10.1051/0004-6361:20041864}, \href
  {https://ui.adsabs.harvard.edu/abs/2005A&A...440..775K} {440, 775}

\bibitem[\protect\citeauthoryear{{Katebi} et~al.,}{{Katebi}
  et~al.}{2019}]{2019MNRAS.487.4057K}
{Katebi} R.,  et~al., 2019, \mn@doi [\mnras] {10.1093/mnras/stz1552}, \href
  {https://ui.adsabs.harvard.edu/abs/2019MNRAS.487.4057K} {487, 4057}

\bibitem[\protect\citeauthoryear{{Kriss} et~al.,}{{Kriss}
  et~al.}{2019}]{2019ApJ...881..153K}
{Kriss} G.~A.,  et~al., 2019, \mn@doi [\apj] {10.3847/1538-4357/ab3049}, \href
  {https://ui.adsabs.harvard.edu/abs/2019ApJ...881..153K} {881, 153}

\bibitem[\protect\citeauthoryear{{LaMassa} et~al.,}{{LaMassa}
  et~al.}{2015}]{2015ApJ...800..144L}
{LaMassa} S.~M.,  et~al., 2015, \mn@doi [\apj] {10.1088/0004-637X/800/2/144},
  \href {https://ui.adsabs.harvard.edu/abs/2015ApJ...800..144L} {800, 144}

\bibitem[\protect\citeauthoryear{{Lefkir}, {Kammoun}, {Barret}, {Boorman},
  {Matzeu}, {Miller}, {Nardini}  \& {Zoghbi}}{{Lefkir}
  et~al.}{2023}]{2023MNRAS.522.1169L}
{Lefkir} M.,  {Kammoun} E.,  {Barret} D.,  {Boorman} P.,  {Matzeu} G.,
  {Miller} J.~M.,  {Nardini} E.,   {Zoghbi} A.,  2023, \mn@doi [\mnras]
  {10.1093/mnras/stad995}, \href
  {https://ui.adsabs.harvard.edu/abs/2023MNRAS.522.1169L} {522, 1169}

\bibitem[\protect\citeauthoryear{{Li} et~al.,}{{Li}
  et~al.}{2022}]{2022ApJ...936...75L}
{Li} S.-S.,  et~al., 2022, \mn@doi [\apj] {10.3847/1538-4357/ac8745}, \href
  {https://ui.adsabs.harvard.edu/abs/2022ApJ...936...75L} {936, 75}

\bibitem[\protect\citeauthoryear{{Liu} \& {Qiao}}{{Liu} \&
  {Qiao}}{2022}]{2022iSci...25j3544L}
{Liu} B.~F.,  {Qiao} E.,  2022, \mn@doi [iScience]
  {10.1016/j.isci.2021.103544}, \href
  {https://ui.adsabs.harvard.edu/abs/2022iSci...25j3544L} {25, 103544}

\bibitem[\protect\citeauthoryear{{Liu}, {Wu}, {Xue}, {Wang}, {Yang}, {Guo}  \&
  {He}}{{Liu} et~al.}{2021}]{2021RAA....21..199L}
{Liu} H.,  {Wu} Q.-W.,  {Xue} Y.-Q.,  {Wang} T.-G.,  {Yang} J.,  {Guo} H.-X.,
  {He} Z.-C.,  2021, \mn@doi [Research in Astronomy and Astrophysics]
  {10.1088/1674-4527/21/8/199}, \href
  {https://ui.adsabs.harvard.edu/abs/2021RAA....21..199L} {21, 199}

\bibitem[\protect\citeauthoryear{{Liu}, {Wu}  \& {Lyu}}{{Liu}
  et~al.}{2022}]{2022ApJ...930...46L}
{Liu} H.,  {Wu} Q.,   {Lyu} B.,  2022, \mn@doi [\apj]
  {10.3847/1538-4357/ac5fa5}, \href
  {https://ui.adsabs.harvard.edu/abs/2022ApJ...930...46L} {930, 46}

\bibitem[\protect\citeauthoryear{{Lyu} \& {Rieke}}{{Lyu} \&
  {Rieke}}{2021}]{2021ApJ...912..126L}
{Lyu} J.,  {Rieke} G.~H.,  2021, \mn@doi [\apj] {10.3847/1538-4357/abee14},
  \href {https://ui.adsabs.harvard.edu/abs/2021ApJ...912..126L} {912, 126}

\bibitem[\protect\citeauthoryear{{Lyu}, {Yan}, {Yu}  \& {Wu}}{{Lyu}
  et~al.}{2021}]{2021MNRAS.506.4188L}
{Lyu} B.,  {Yan} Z.,  {Yu} W.,   {Wu} Q.,  2021, \mn@doi [\mnras]
  {10.1093/mnras/stab1581}, \href
  {https://ui.adsabs.harvard.edu/abs/2021MNRAS.506.4188L} {506, 4188}

\bibitem[\protect\citeauthoryear{{Lyu}, {Wu}, {Yan}, {Yu}  \& {Liu}}{{Lyu}
  et~al.}{2022}]{2022ApJ...927..227L}
{Lyu} B.,  {Wu} Q.,  {Yan} Z.,  {Yu} W.,   {Liu} H.,  2022, \mn@doi [\apj]
  {10.3847/1538-4357/ac5256}, \href
  {https://ui.adsabs.harvard.edu/abs/2022ApJ...927..227L} {927, 227}

\bibitem[\protect\citeauthoryear{{MacLeod} et~al.,}{{MacLeod}
  et~al.}{2019}]{2019ApJ...874....8M}
{MacLeod} C.~L.,  et~al., 2019, \mn@doi [\apj] {10.3847/1538-4357/ab05e2},
  \href {https://ui.adsabs.harvard.edu/abs/2019ApJ...874....8M} {874, 8}

\bibitem[\protect\citeauthoryear{{Magdziarz} \& {Zdziarski}}{{Magdziarz} \&
  {Zdziarski}}{1995}]{1995MNRAS.273..837M}
{Magdziarz} P.,  {Zdziarski} A.~A.,  1995, \mn@doi [\mnras]
  {10.1093/mnras/273.3.837}, \href
  {https://ui.adsabs.harvard.edu/abs/1995MNRAS.273..837M} {273, 837}

\bibitem[\protect\citeauthoryear{{Marchesi}, {Ajello}, {Marcotulli},
  {Comastri}, {Lanzuisi}  \& {Vignali}}{{Marchesi}
  et~al.}{2018}]{2018ApJ...854...49M}
{Marchesi} S.,  {Ajello} M.,  {Marcotulli} L.,  {Comastri} A.,  {Lanzuisi} G.,
   {Vignali} C.,  2018, \mn@doi [\apj] {10.3847/1538-4357/aaa410}, \href
  {https://ui.adsabs.harvard.edu/abs/2018ApJ...854...49M} {854, 49}

\bibitem[\protect\citeauthoryear{{Marchesi} et~al.,}{{Marchesi}
  et~al.}{2022}]{2022ApJ...935..114M}
{Marchesi} S.,  et~al., 2022, \mn@doi [\apj] {10.3847/1538-4357/ac80be}, \href
  {https://ui.adsabs.harvard.edu/abs/2022ApJ...935..114M} {935, 114}

\bibitem[\protect\citeauthoryear{{Marin}, {Porquet}, {Goosmann},
  {Dov{\v{c}}iak}, {Muleri}, {Grosso}  \& {Karas}}{{Marin}
  et~al.}{2013}]{2013MNRAS.436.1615M}
{Marin} F.,  {Porquet} D.,  {Goosmann} R.~W.,  {Dov{\v{c}}iak} M.,  {Muleri}
  F.,  {Grosso} N.,   {Karas} V.,  2013, \mn@doi [\mnras]
  {10.1093/mnras/stt1677}, \href
  {https://ui.adsabs.harvard.edu/abs/2013MNRAS.436.1615M} {436, 1615}

\bibitem[\protect\citeauthoryear{{Marin}, {Hutsem{\'e}kers}  \& {Ag{\'\i}s
  Gonz{\'a}lez}}{{Marin} et~al.}{2019}]{2019sf2a.conf..509M}
{Marin} F.,  {Hutsem{\'e}kers} D.,   {Ag{\'\i}s Gonz{\'a}lez} B.,  2019, in {Di
  Matteo} P.,  {Creevey} O.,  {Crida} A.,  {Kordopatis} G.,  {Malzac} J.,
  {Marquette} J.~B.,  {N'Diaye} M.,   {Venot} O.,  eds, SF2A-2019: Proceedings
  of the Annual meeting of the French Society of Astronomy and Astrophysics.
  p.~Di (\mn@eprint {arXiv} {1909.02801})

\bibitem[\protect\citeauthoryear{{Matt}}{{Matt}}{2002}]{2002RSPTA.360.2045M}
{Matt} G.,  2002, \mn@doi [Philosophical Transactions of the Royal Society of
  London Series A] {10.1098/rsta.2002.1052}, \href
  {https://ui.adsabs.harvard.edu/abs/2002RSPTA.360.2045M} {360, 2045}

\bibitem[\protect\citeauthoryear{{Matt}, {Guainazzi}  \& {Maiolino}}{{Matt}
  et~al.}{2003}]{2003MNRAS.342..422M}
{Matt} G.,  {Guainazzi} M.,   {Maiolino} R.,  2003, \mn@doi [\mnras]
  {10.1046/j.1365-8711.2003.06539.x}, \href
  {https://ui.adsabs.harvard.edu/abs/2003MNRAS.342..422M} {342, 422}

\bibitem[\protect\citeauthoryear{{Matthews}, {Knigge}, {Higginbottom}, {Long},
  {Sim}, {Mangham}, {Parkinson}  \& {Hewitt}}{{Matthews}
  et~al.}{2020}]{2020MNRAS.492.5540M}
{Matthews} J.~H.,  {Knigge} C.,  {Higginbottom} N.,  {Long} K.~S.,  {Sim}
  S.~A.,  {Mangham} S.~W.,  {Parkinson} E.~J.,   {Hewitt} H.~A.,  2020, \mn@doi
  [\mnras] {10.1093/mnras/staa136}, \href
  {https://ui.adsabs.harvard.edu/abs/2020MNRAS.492.5540M} {492, 5540}

\bibitem[\protect\citeauthoryear{{Mehdipour}, {Kriss}, {Kaastra}, {Costantini},
  {Gu}, {Landt}, {Mao}  \& {Rogantini}}{{Mehdipour}
  et~al.}{2024}]{2024ApJ...962..155M}
{Mehdipour} M.,  {Kriss} G.~A.,  {Kaastra} J.~S.,  {Costantini} E.,  {Gu} L.,
  {Landt} H.,  {Mao} J.,   {Rogantini} D.,  2024, \mn@doi [\apj]
  {10.3847/1538-4357/ad1bcb}, \href
  {https://ui.adsabs.harvard.edu/abs/2024ApJ...962..155M} {962, 155}

\bibitem[\protect\citeauthoryear{{Mondal}, {Adhikari}, {Hryniewicz}, {Stalin}
  \& {Pandey}}{{Mondal} et~al.}{2022}]{2022A&A...662A..77M}
{Mondal} S.,  {Adhikari} T.~P.,  {Hryniewicz} K.,  {Stalin} C.~S.,   {Pandey}
  A.,  2022, \mn@doi [\aap] {10.1051/0004-6361/202243084}, \href
  {https://ui.adsabs.harvard.edu/abs/2022A&A...662A..77M} {662, A77}

\bibitem[\protect\citeauthoryear{{Netzer}}{{Netzer}}{2015}]{Netzer2015}
{Netzer} H.,  2015, \mn@doi [\araa] {10.1146/annurev-astro-082214-122302},
  \href {https://ui.adsabs.harvard.edu/abs/2015ARA&A..53..365N} {53, 365}

\bibitem[\protect\citeauthoryear{{Netzer}}{{Netzer}}{2019}]{2019MNRAS.488.5185N}
{Netzer} H.,  2019, \mn@doi [\mnras] {10.1093/mnras/stz2016}, \href
  {https://ui.adsabs.harvard.edu/abs/2019MNRAS.488.5185N} {488, 5185}

\bibitem[\protect\citeauthoryear{{Neustadt} et~al.,}{{Neustadt}
  et~al.}{2023}]{2023MNRAS.tmp..710N}
{Neustadt} J.~M.~M.,  et~al., 2023, \mn@doi [\mnras] {10.1093/mnras/stad725},
  \href {https://ui.adsabs.harvard.edu/abs/2023MNRAS.tmp..710N} {}

\bibitem[\protect\citeauthoryear{{Noda} \& {Done}}{{Noda} \&
  {Done}}{2018}]{2018MNRAS.480.3898N}
{Noda} H.,  {Done} C.,  2018, \mn@doi [\mnras] {10.1093/mnras/sty2032}, \href
  {https://ui.adsabs.harvard.edu/abs/2018MNRAS.480.3898N} {480, 3898}

\bibitem[\protect\citeauthoryear{{Noda} et~al.,}{{Noda}
  et~al.}{2022}]{2022arXiv221202731N}
{Noda} H.,  et~al., 2022, arXiv e-prints, \href
  {https://ui.adsabs.harvard.edu/abs/2022arXiv221202731N} {p. arXiv:2212.02731}

\bibitem[\protect\citeauthoryear{{Oh} et~al.,}{{Oh}
  et~al.}{2022}]{2022ApJS..261....4O}
{Oh} K.,  et~al., 2022, \mn@doi [\apjs] {10.3847/1538-4365/ac5b68}, \href
  {https://ui.adsabs.harvard.edu/abs/2022ApJS..261....4O} {261, 4}

\bibitem[\protect\citeauthoryear{{Oknyanskij}, {Metlova}, {Huseynov}, {Guo}  \&
  {Lyuty}}{{Oknyanskij} et~al.}{2016}]{2016OAP....29...95O}
{Oknyanskij} V.~L.,  {Metlova} N.~V.,  {Huseynov} N.~A.,  {Guo} D.-F.,
  {Lyuty} V.~M.,  2016, \mn@doi [Odessa Astronomical Publications]
  {10.18524/1810-4215.2016.29.85058}, \href
  {https://ui.adsabs.harvard.edu/abs/2016OAP....29...95O} {29, 95}

\bibitem[\protect\citeauthoryear{{Oknyansky}}{{Oknyansky}}{2022}]{2022AN....34310080O}
{Oknyansky} V.,  2022, \mn@doi [Astronomische Nachrichten]
  {10.1002/asna.20210080}, \href
  {https://ui.adsabs.harvard.edu/abs/2022AN....34310080O} {343, e210080}

\bibitem[\protect\citeauthoryear{{Oknyansky}, {Gaskell}  \&
  {Shimanovskaya}}{{Oknyansky} et~al.}{2015}]{2015OAP....28..175O}
{Oknyansky} V.~L.,  {Gaskell} C.~M.,   {Shimanovskaya} E.~V.,  2015, \mn@doi
  [Odessa Astronomical Publications] {10.48550/arXiv.1511.02170}, \href
  {https://ui.adsabs.harvard.edu/abs/2015OAP....28..175O} {28, 175}

\bibitem[\protect\citeauthoryear{{Oknyansky}, {Winkler}, {Tsygankov},
  {Lipunov}, {Gorbovskoy}, {van Wyk}, {Buckley}  \& {Tyurina}}{{Oknyansky}
  et~al.}{2019}]{2019MNRAS.483..558O}
{Oknyansky} V.~L.,  {Winkler} H.,  {Tsygankov} S.~S.,  {Lipunov} V.~M.,
  {Gorbovskoy} E.~S.,  {van Wyk} F.,  {Buckley} D.~A.~H.,   {Tyurina} N.~V.,
  2019, \mn@doi [\mnras] {10.1093/mnras/sty3133}, \href
  {https://ui.adsabs.harvard.edu/abs/2019MNRAS.483..558O} {483, 558}

\bibitem[\protect\citeauthoryear{{Onori} et~al.,}{{Onori}
  et~al.}{2017}]{2017MNRAS.468L..97O}
{Onori} F.,  et~al., 2017, \mn@doi [\mnras] {10.1093/mnrasl/slx032}, \href
  {https://ui.adsabs.harvard.edu/abs/2017MNRAS.468L..97O} {468, L97}

\bibitem[\protect\citeauthoryear{{Osterbrock}}{{Osterbrock}}{1981}]{1981ApJ...249..462O}
{Osterbrock} D.~E.,  1981, \mn@doi [\apj] {10.1086/159306}, \href
  {https://ui.adsabs.harvard.edu/abs/1981ApJ...249..462O} {249, 462}

\bibitem[\protect\citeauthoryear{{Osterbrock} \& {Koski}}{{Osterbrock} \&
  {Koski}}{1976}]{1976MNRAS.176P..61O}
{Osterbrock} D.~E.,  {Koski} A.~T.,  1976, \mn@doi [\mnras]
  {10.1093/mnras/176.1.61P}, \href
  {https://ui.adsabs.harvard.edu/abs/1976MNRAS.176P..61O} {176, 61P}

\bibitem[\protect\citeauthoryear{{Pal}, {Stalin}, {Mallick}  \& {Rani}}{{Pal}
  et~al.}{2022}]{Pal2022}
{Pal} I.,  {Stalin} C.~S.,  {Mallick} L.,   {Rani} P.,  2022, \mn@doi [\aap]
  {10.1051/0004-6361/202142386}, \href
  {https://ui.adsabs.harvard.edu/abs/2022A&A...662A..78P} {662, A78}

\bibitem[\protect\citeauthoryear{{Pancoast}, {Brewer}, {Treu}, {Park}, {Barth},
  {Bentz}  \& {Woo}}{{Pancoast} et~al.}{2014}]{2014MNRAS.445.3073P}
{Pancoast} A.,  {Brewer} B.~J.,  {Treu} T.,  {Park} D.,  {Barth} A.~J.,
  {Bentz} M.~C.,   {Woo} J.-H.,  2014, \mn@doi [\mnras]
  {10.1093/mnras/stu1419}, \href
  {https://ui.adsabs.harvard.edu/abs/2014MNRAS.445.3073P} {445, 3073}

\bibitem[\protect\citeauthoryear{{Panda} \& {{\'S}niegowska}}{{Panda} \&
  {{\'S}niegowska}}{2024}]{2024ApJS..272...13P}
{Panda} S.,  {{\'S}niegowska} M.,  2024, \mn@doi [\apjs]
  {10.3847/1538-4365/ad344f}, \href
  {https://ui.adsabs.harvard.edu/abs/2024ApJS..272...13P} {272, 13}

\bibitem[\protect\citeauthoryear{{Paturel}, {Petit}, {Prugniel}, {Theureau},
  {Rousseau}, {Brouty}, {Dubois}  \& {Cambr{\'e}sy}}{{Paturel}
  et~al.}{2003}]{Paturel2003}
{Paturel} G.,  {Petit} C.,  {Prugniel} P.,  {Theureau} G.,  {Rousseau} J.,
  {Brouty} M.,  {Dubois} P.,   {Cambr{\'e}sy} L.,  2003, \mn@doi [\aap]
  {10.1051/0004-6361:20031411}, \href
  {https://ui.adsabs.harvard.edu/abs/2003A&A...412...45P} {412, 45}

\bibitem[\protect\citeauthoryear{{Piconcelli}, {Bianchi}, {Guainazzi}, {Fiore}
  \& {Chiaberge}}{{Piconcelli} et~al.}{2007}]{Piconcelli2007}
{Piconcelli} E.,  {Bianchi} S.,  {Guainazzi} M.,  {Fiore} F.,   {Chiaberge} M.,
   2007, \mn@doi [\aap] {10.1051/0004-6361:20066439}, \href
  {https://ui.adsabs.harvard.edu/abs/2007A&A...466..855P} {466, 855}

\bibitem[\protect\citeauthoryear{{Potts} \& {Villforth}}{{Potts} \&
  {Villforth}}{2021}]{2021A&A...650A..33P}
{Potts} B.,  {Villforth} C.,  2021, \mn@doi [\aap]
  {10.1051/0004-6361/202140597}, \href
  {https://ui.adsabs.harvard.edu/abs/2021A&A...650A..33P} {650, A33}

\bibitem[\protect\citeauthoryear{{Puccetti}, {Fiore}, {Risaliti}, {Capalbi},
  {Elvis}  \& {Nicastro}}{{Puccetti} et~al.}{2007}]{2007MNRAS.377..607P}
{Puccetti} S.,  {Fiore} F.,  {Risaliti} G.,  {Capalbi} M.,  {Elvis} M.,
  {Nicastro} F.,  2007, \mn@doi [\mnras] {10.1111/j.1365-2966.2007.11634.x},
  \href {https://ui.adsabs.harvard.edu/abs/2007MNRAS.377..607P} {377, 607}

\bibitem[\protect\citeauthoryear{{Qiao} \& {Liu}}{{Qiao} \&
  {Liu}}{2013}]{2013ApJ...764....2Q}
{Qiao} E.,  {Liu} B.~F.,  2013, \mn@doi [\apj] {10.1088/0004-637X/764/1/2},
  \href {https://ui.adsabs.harvard.edu/abs/2013ApJ...764....2Q} {764, 2}

\bibitem[\protect\citeauthoryear{{Raimundo}, {Vestergaard}, {Koay}, {Lawther},
  {Casasola}  \& {Peterson}}{{Raimundo} et~al.}{2019}]{2019MNRAS.486..123R}
{Raimundo} S.~I.,  {Vestergaard} M.,  {Koay} J.~Y.,  {Lawther} D.,  {Casasola}
  V.,   {Peterson} B.~M.,  2019, \mn@doi [\mnras] {10.1093/mnras/stz852}, \href
  {https://ui.adsabs.harvard.edu/abs/2019MNRAS.486..123R} {486, 123}

\bibitem[\protect\citeauthoryear{{Ricci} \& {Trakhtenbrot}}{{Ricci} \&
  {Trakhtenbrot}}{2023}]{2023NatAs...7.1282R}
{Ricci} C.,  {Trakhtenbrot} B.,  2023, \mn@doi [Nature Astronomy]
  {10.1038/s41550-023-02108-4}, \href
  {https://ui.adsabs.harvard.edu/abs/2023NatAs...7.1282R} {7, 1282}

\bibitem[\protect\citeauthoryear{{Ricci} et~al.,}{{Ricci}
  et~al.}{2016}]{2016ApJ...820....5R}
{Ricci} C.,  et~al., 2016, \mn@doi [\apj] {10.3847/0004-637X/820/1/5}, \href
  {https://ui.adsabs.harvard.edu/abs/2016ApJ...820....5R} {820, 5}

\bibitem[\protect\citeauthoryear{{Ricci} et~al.,}{{Ricci}
  et~al.}{2017}]{2017Natur.549..488R}
{Ricci} C.,  et~al., 2017, \mn@doi [\nat] {10.1038/nature23906}, \href
  {https://ui.adsabs.harvard.edu/abs/2017Natur.549..488R} {549, 488}

\bibitem[\protect\citeauthoryear{{Ricci}, {Steiner}, {May}, {Garcia-Rissmann}
  \& {Menezes}}{{Ricci} et~al.}{2018a}]{2018MNRAS.473.5334R}
{Ricci} T.~V.,  {Steiner} J.~E.,  {May} D.,  {Garcia-Rissmann} A.,   {Menezes}
  R.~B.,  2018a, \mn@doi [\mnras] {10.1093/mnras/stx2746}, \href
  {https://ui.adsabs.harvard.edu/abs/2018MNRAS.473.5334R} {473, 5334}

\bibitem[\protect\citeauthoryear{{Ricci} et~al.,}{{Ricci}
  et~al.}{2018b}]{2018MNRAS.480.1819R}
{Ricci} C.,  et~al., 2018b, \mn@doi [\mnras] {10.1093/mnras/sty1879}, \href
  {https://ui.adsabs.harvard.edu/abs/2018MNRAS.480.1819R} {480, 1819}

\bibitem[\protect\citeauthoryear{{Risaliti}, {Elvis}, {Fabbiano}, {Baldi}  \&
  {Zezas}}{{Risaliti} et~al.}{2005}]{2005ApJ...623L..93R}
{Risaliti} G.,  {Elvis} M.,  {Fabbiano} G.,  {Baldi} A.,   {Zezas} A.,  2005,
  \mn@doi [\apjl] {10.1086/430252}, \href
  {https://ui.adsabs.harvard.edu/abs/2005ApJ...623L..93R} {623, L93}

\bibitem[\protect\citeauthoryear{{Risaliti} et~al.,}{{Risaliti}
  et~al.}{2009}]{2009MNRAS.393L...1R}
{Risaliti} G.,  et~al., 2009, \mn@doi [\mnras]
  {10.1111/j.1745-3933.2008.00580.x}, \href
  {https://ui.adsabs.harvard.edu/abs/2009MNRAS.393L...1R} {393, L1}

\bibitem[\protect\citeauthoryear{{Rivers} et~al.,}{{Rivers}
  et~al.}{2015}]{2015ApJ...815...55R}
{Rivers} E.,  et~al., 2015, \mn@doi [\apj] {10.1088/0004-637X/815/1/55}, \href
  {https://ui.adsabs.harvard.edu/abs/2015ApJ...815...55R} {815, 55}

\bibitem[\protect\citeauthoryear{{Senarath} et~al.,}{{Senarath}
  et~al.}{2021}]{2021MNRAS.503.2583S}
{Senarath} M.~R.,  et~al., 2021, \mn@doi [\mnras] {10.1093/mnras/stab393},
  \href {https://ui.adsabs.harvard.edu/abs/2021MNRAS.503.2583S} {503, 2583}

\bibitem[\protect\citeauthoryear{{Shakura} \& {Sunyaev}}{{Shakura} \&
  {Sunyaev}}{1973}]{1973A&A....24..337S}
{Shakura} N.~I.,  {Sunyaev} R.~A.,  1973, \aap, \href
  {https://ui.adsabs.harvard.edu/abs/1973A&A....24..337S} {24, 337}

\bibitem[\protect\citeauthoryear{{Shappee} et~al.,}{{Shappee}
  et~al.}{2014}]{2014ApJ...788...48S}
{Shappee} B.~J.,  et~al., 2014, \mn@doi [\apj] {10.1088/0004-637X/788/1/48},
  \href {https://ui.adsabs.harvard.edu/abs/2014ApJ...788...48S} {788, 48}

\bibitem[\protect\citeauthoryear{{Sheng}, {Wang}, {Jiang}, {Yang}, {Yan}, {Dou}
   \& {Peng}}{{Sheng} et~al.}{2017}]{2017ApJ...846L...7S}
{Sheng} Z.,  {Wang} T.,  {Jiang} N.,  {Yang} C.,  {Yan} L.,  {Dou} L.,   {Peng}
  B.,  2017, \mn@doi [\apjl] {10.3847/2041-8213/aa85de}, \href
  {https://ui.adsabs.harvard.edu/abs/2017ApJ...846L...7S} {846, L7}

\bibitem[\protect\citeauthoryear{{Sheng} et~al.,}{{Sheng}
  et~al.}{2020}]{2020ApJ...889...46S}
{Sheng} Z.,  et~al., 2020, \mn@doi [\apj] {10.3847/1538-4357/ab5af9}, \href
  {https://ui.adsabs.harvard.edu/abs/2020ApJ...889...46S} {889, 46}

\bibitem[\protect\citeauthoryear{{Sniegowska}, {Czerny}, {Bon}  \&
  {Bon}}{{Sniegowska} et~al.}{2020}]{2020A&A...641A.167S}
{Sniegowska} M.,  {Czerny} B.,  {Bon} E.,   {Bon} N.,  2020, \mn@doi [\aap]
  {10.1051/0004-6361/202038575}, \href
  {https://ui.adsabs.harvard.edu/abs/2020A&A...641A.167S} {641, A167}

\bibitem[\protect\citeauthoryear{{Tanimoto}, {Ueda}, {Odaka}, {Kawaguchi},
  {Fukazawa}  \& {Kawamuro}}{{Tanimoto} et~al.}{2019}]{2019ApJ...877...95T}
{Tanimoto} A.,  {Ueda} Y.,  {Odaka} H.,  {Kawaguchi} T.,  {Fukazawa} Y.,
  {Kawamuro} T.,  2019, \mn@doi [\apj] {10.3847/1538-4357/ab1b20}, \href
  {https://ui.adsabs.harvard.edu/abs/2019ApJ...877...95T} {877, 95}

\bibitem[\protect\citeauthoryear{{Tanimoto}, {Ueda}, {Odaka}, {Yamada}  \&
  {Ricci}}{{Tanimoto} et~al.}{2022}]{2022ApJS..260...30T}
{Tanimoto} A.,  {Ueda} Y.,  {Odaka} H.,  {Yamada} S.,   {Ricci} C.,  2022,
  \mn@doi [\apjs] {10.3847/1538-4365/ac5f59}, \href
  {https://ui.adsabs.harvard.edu/abs/2022ApJS..260...30T} {260, 30}

\bibitem[\protect\citeauthoryear{{Temple} et~al.,}{{Temple}
  et~al.}{2023}]{2023MNRAS.518.2938T}
{Temple} M.~J.,  et~al., 2023, \mn@doi [\mnras] {10.1093/mnras/stac3279}, \href
  {https://ui.adsabs.harvard.edu/abs/2023MNRAS.518.2938T} {518, 2938}

\bibitem[\protect\citeauthoryear{{Trakhtenbrot} et~al.,}{{Trakhtenbrot}
  et~al.}{2017}]{2017MNRAS.470..800T}
{Trakhtenbrot} B.,  et~al., 2017, \mn@doi [\mnras] {10.1093/mnras/stx1117},
  \href {https://ui.adsabs.harvard.edu/abs/2017MNRAS.470..800T} {470, 800}

\bibitem[\protect\citeauthoryear{{Trakhtenbrot} et~al.,}{{Trakhtenbrot}
  et~al.}{2019}]{2019ApJ...883...94T}
{Trakhtenbrot} B.,  et~al., 2019, \mn@doi [\apj] {10.3847/1538-4357/ab39e4},
  \href {https://ui.adsabs.harvard.edu/abs/2019ApJ...883...94T} {883, 94}

\bibitem[\protect\citeauthoryear{{Turner}, {Reeves}, {Braito}, {Lobban},
  {Kraemer}  \& {Miller}}{{Turner} et~al.}{2018}]{2018MNRAS.481.2470T}
{Turner} T.~J.,  {Reeves} J.~N.,  {Braito} V.,  {Lobban} A.,  {Kraemer} S.,
  {Miller} L.,  2018, \mn@doi [\mnras] {10.1093/mnras/sty2447}, \href
  {https://ui.adsabs.harvard.edu/abs/2018MNRAS.481.2470T} {481, 2470}

\bibitem[\protect\citeauthoryear{{Walton}, {Reis}  \& {Fabian}}{{Walton}
  et~al.}{2010}]{2010MNRAS.408..601W}
{Walton} D.~J.,  {Reis} R.~C.,   {Fabian} A.~C.,  2010, \mn@doi [\mnras]
  {10.1111/j.1365-2966.2010.17148.x}, \href
  {https://ui.adsabs.harvard.edu/abs/2010MNRAS.408..601W} {408, 601}

\bibitem[\protect\citeauthoryear{{Wang}, {Xu}, {Wang}, {Zhang}, {Zheng}  \&
  {Wei}}{{Wang} et~al.}{2019}]{2019ApJ...887...15W}
{Wang} J.,  {Xu} D.~W.,  {Wang} Y.,  {Zhang} J.~B.,  {Zheng} J.,   {Wei} J.~Y.,
   2019, \mn@doi [\apj] {10.3847/1538-4357/ab4d90}, \href
  {https://ui.adsabs.harvard.edu/abs/2019ApJ...887...15W} {887, 15}

\bibitem[\protect\citeauthoryear{{Wang}, {Xu}  \& {Wei}}{{Wang}
  et~al.}{2020}]{2020ApJ...901....1W}
{Wang} J.,  {Xu} D.~W.,   {Wei} J.~Y.,  2020, \mn@doi [\apj]
  {10.3847/1538-4357/abaa48}, \href
  {https://ui.adsabs.harvard.edu/abs/2020ApJ...901....1W} {901, 1}

\bibitem[\protect\citeauthoryear{{Wang}, {Zheng}, {Xu}, {Brink}, {Filippenko},
  {Gao}, {Sun}  \& {Wei}}{{Wang} et~al.}{2022}]{2022RAA....22a5011W}
{Wang} J.,  {Zheng} W.~K.,  {Xu} D.~W.,  {Brink} T.~G.,  {Filippenko} A.~V.,
  {Gao} C.,  {Sun} S.~S.,   {Wei} J.~Y.,  2022, \mn@doi [Research in Astronomy
  and Astrophysics] {10.1088/1674-4527/ac3477}, \href
  {https://ui.adsabs.harvard.edu/abs/2022RAA....22a5011W} {22, 015011}

\bibitem[\protect\citeauthoryear{{Wold}, {Lacy}, {K{\"a}ufl}  \&
  {Siebenmorgen}}{{Wold} et~al.}{2006}]{2006A&A...460..449W}
{Wold} M.,  {Lacy} M.,  {K{\"a}ufl} H.~U.,   {Siebenmorgen} R.,  2006, \mn@doi
  [\aap] {10.1051/0004-6361:20053385}, \href
  {https://ui.adsabs.harvard.edu/abs/2006A&A...460..449W} {460, 449}

\bibitem[\protect\citeauthoryear{{Yang} et~al.,}{{Yang}
  et~al.}{2018}]{2018ApJ...862..109Y}
{Yang} Q.,  et~al., 2018, \mn@doi [\apj] {10.3847/1538-4357/aaca3a}, \href
  {https://ui.adsabs.harvard.edu/abs/2018ApJ...862..109Y} {862, 109}

\bibitem[\protect\citeauthoryear{{Yang} et~al.,}{{Yang}
  et~al.}{2023}]{2023ApJ...953...61Y}
{Yang} Q.,  et~al., 2023, \mn@doi [\apj] {10.3847/1538-4357/acdedd}, \href
  {https://ui.adsabs.harvard.edu/abs/2023ApJ...953...61Y} {953, 61}

\bibitem[\protect\citeauthoryear{{Yaqoob}, {Warwick}  \& {Pounds}}{{Yaqoob}
  et~al.}{1989}]{1989MNRAS.236..153Y}
{Yaqoob} T.,  {Warwick} R.~S.,   {Pounds} K.~A.,  1989, \mn@doi [\mnras]
  {10.1093/mnras/236.2.153}, \href
  {https://ui.adsabs.harvard.edu/abs/1989MNRAS.236..153Y} {236, 153}

\bibitem[\protect\citeauthoryear{{Yuan} \& {Narayan}}{{Yuan} \&
  {Narayan}}{2014}]{2014ARA&A..52..529Y}
{Yuan} F.,  {Narayan} R.,  2014, \mn@doi [\araa]
  {10.1146/annurev-astro-082812-141003}, \href
  {https://ui.adsabs.harvard.edu/abs/2014ARA&A..52..529Y} {52, 529}

\bibitem[\protect\citeauthoryear{{Zeltyn} et~al.,}{{Zeltyn}
  et~al.}{2022}]{2022ApJ...939L..16Z}
{Zeltyn} G.,  et~al., 2022, \mn@doi [\apjl] {10.3847/2041-8213/ac9a47}, \href
  {https://ui.adsabs.harvard.edu/abs/2022ApJ...939L..16Z} {939, L16}

\bibitem[\protect\citeauthoryear{{Zeltyn} et~al.,}{{Zeltyn}
  et~al.}{2024}]{2024ApJ...966...85Z}
{Zeltyn} G.,  et~al., 2024, \mn@doi [\apj] {10.3847/1538-4357/ad2f30}, \href
  {https://ui.adsabs.harvard.edu/abs/2024ApJ...966...85Z} {966, 85}

\bibitem[\protect\citeauthoryear{{Zhong} \& {Wang}}{{Zhong} \&
  {Wang}}{2022}]{2022RAA....22c5002Z}
{Zhong} X.-G.,  {Wang} J.-C.,  2022, \mn@doi [Research in Astronomy and
  Astrophysics] {10.1088/1674-4527/ac42c0}, \href
  {https://ui.adsabs.harvard.edu/abs/2022RAA....22c5002Z} {22, 035002}

\makeatother
\end{thebibliography}




\appendix
\section{Sample}
\subsection{ESO 362-G18}
ESO 362-G18 is a nearby (z = 0.0124) Seyfert galaxy with BH mass log($M_\mathrm{BH}/M_{\odot})$ = $  7.65$ and a high black hole spin $a \ge 0.92$ \citep[see ][and references therein]{2014MNRAS.443.2862A,2018Galax...6...52A}. An inclination ($i$ = $ 53^{\circ} \pm 5^{\circ}$) between the disc axis and our line-of-sight is well constrained through a disc-reflection component accounting for the broad Fe K$\alpha$ line and hard X-ray band above 10 keV \citep[][]{2000ApJ...532..867F,2014MNRAS.443.2862A}.  

ESO 362-G18 was classified as Seyfert (Sy) 1.5, however, experienced multiple times of variations of broad emission lines. Based on four spectra of different epochs, \citet{2018Galax...6...52A} reported the Sy1.9 classification in 6dF (30 Jan 2003) and EFOSC1 (21 Sep 2006) data and Sy1.5 classification with evident broad emission lines (especially for $\rm H\beta$ lines) in EMMI (18 Sep 2004) and FORS2 (29 Mar 2016) data. 

ESO 362-G18 also showed highly variable column density with a mild absorption ($N_\mathrm{H} \sim 0.5 \times\,10^{22}\, \mathrm{cm}^{-2}$) in \swift\, data (26 Nov 2005). A more heavy absorption ($N_\mathrm{H} \sim 3-4 \times\,10^{23}\, \mathrm{cm}^{-2}$) by two orders of magnitude was found after 2 months (28 Jan 2006) in \xmm\, data. The absorption returned to $N_\mathrm{H}  \lesssim 3 \times\,10^{22}\, \mathrm{cm}^{-2}$ in \suzaku\,, \xmm\,, and \chandra\, data during 2008--2010 \citep[][]{2014MNRAS.443.2862A}.

\subsection{NGC 1365}

NGC 1365 is nearby (z = 0.0055) Seyfert galaxy with a BH mass log($M_\mathrm{BH}/M_{\odot}) $=$ 6.65$ \citep{2017MNRAS.468L..97O}. Based on the \suzaku\, data, the inclination angle of the accretion disc ($i $=$ 57^{\circ +3^{\circ}}_{-2^{\circ}} $) was constrained with a disc reflection component \citep{2010MNRAS.408..601W}. NGC 1365 was a well-known Changing-obscuration AGN with a rapid X-ray spectral state change from Compton-thick to Compton-thin and back again in 2002 and 2003. Significant variability of hydrogen column density and strong spectral evolution in optical and X-ray were also found between 2012 and 2013 based on \xmm\, and \nustar\, data \citep{2021RAA....21..199L}. 
NGC 1365 also experienced multiple times of rapid variations of broad Balmer lines during the past years. NGC 1365 was originally observed to show broad Balmer lines in 1993 August and then turned into a faint state with only narrow $\rm H\beta$ and weak broad $\rm H\alpha$ lines in 2009 January and 2010 September. It turned back on with broad Balmer lines in 2013 January, 2013 December, 2014 October, and 2017 June spectra and returned to the turning-off state with only narrow lines newly discovered in 2021 December \citep[see ][and references therein]{2023MNRAS.518.2938T}. 


\subsection{NGC 4151}
NGC 4151 is a well-studied nearby (z = 0.0033) changing-look Seyfert galaxy with BH mass log($M_\mathrm{BH}/M_{\odot}$) around 7.56 \citep[e.g.][]{2013ApJ...773...90G}. A thick disc with an inclination ($i \sim 58.1^{\circ +8.4^{\circ}}_{-9.6^{\circ}} $) was estimated \citep[][]{2022ApJ...934..168B}, which is consistent with constraints from geometric modeling of the narrow-line region in previous works and the BLR gas cloud temporarily obscured the central X-ray emission \citep[see details in][]{2022ApJ...934..168B}. 

NGC 4151 experienced a rapid change of $N_\mathrm{H} \sim 10^{22}-10^{23}\, \mathrm{cm}^{-2}$ in 1996 July and 2001 Dec \citep[see][]{2007MNRAS.377..607P}. NGC 4151 also experienced strong historic variability \citep[e.g.,][]{2016OAP....29...95O} and multiple optical spectral type changes from Sy1 in 1974 to Sy1.9 during 1984-1989 and to Sy1.5 during 1990-1998 and to Sy1.8 in 2001 \citep[e.g.,][]{2019sf2a.conf..509M}. The broad \hb\, lines are visible between 2005 February 27 and 2005 April 10 in a reverberation mapping program \citep[][]{2022ApJ...934..168B}. The long-term reverberation mapping campaigns of NGC 4151 also show the broad \hb\, lines since 2018 \citep[e.g.,][]{2022ApJ...936...75L,2023MNRAS.520.1807C}.  

\subsection{NGC 5548}
NGC 5548 is a well-known Seyfert 1 galaxy with BH mass log($M_\mathrm{BH}/M_{\odot}$) around 7.7 \citep[e.g.][]{2015PASP..127...67B}. A narrow thick disk with an inclination angle  ($i $=$ 38.8^{\circ +12.1^{\circ}}_{-11.4^{\circ}} $) is constrained for the geometry of \hb \,BLR \citep{2014MNRAS.445.3073P}. NGC 5548 showed a clear changing-obscuration event in 2013 with two obscuring components, which lasted several years \citep{2014Sci...345...64K,2023NatAs...7.1282R}. NGC 5548 also showed variations of both the intrinsic continuum and the obscurer \citep{2015A&A...579A..42D,2019ApJ...881..153K,2020A&A...641A.167S,2024ApJ...962..155M}. NGC 5548 experienced transitions from type 1.0 (1978--2001) to type 1.8 (2005--2007) and to 1.0 \citep[2014--2021; see][]{2024arXiv241108676J}.

\subsection{NGC 7582}
NGC 7582 is a nearby (z = 0.0053) changing-look Seyfert galaxy with BH mass log($M_\mathrm{BH}/M_{\odot}) $=$ 7.74$ \citep{2018MNRAS.473.5334R} and an galactic inclination angle of ($i= 68^{\circ} $) \citep[e.g.][]{Paturel2003}, which exhibits complex and variable X-ray spectra. A lower limit for the disk inclination angle $i \sim 70.8^{\circ +10.6^{\circ} } $ was constrained for NGC 7582 through X-ray spectral fitting based on the \nustar\, data combined with the simultaneous \xmm\, data \citep[see ][]{2022ApJS..260...30T}. An inclination angle ($i = 80^{\circ}$) of the line of sight to the disk was used to fit the X-ray spectra of NGC 7582 as a Seyfert 2\citep{2023MNRAS.522.1169L}.

\citet{Piconcelli2007} found that both the X-ray spectral index ($\Gamma$) and $N_\mathrm{H,los}$ are variable with \xmm\, data. Rapid variations of column density from an inner absorber ($N_\mathrm{H,i}$ ranged between 3.3 -- 12  $\times\,10^{23}\, \mathrm{cm}^{-2}$ down to timescales of less than a day) was monitored by \suzaku\, and \xmm\, during 2007 \citep{2009ApJ...695..781B}. NGC 7582 became more heavily absorbed $N_\mathrm{H,los}\sim (3.1\pm 0.7)\times 10^{23}\, \mathrm{cm}^{-2} $ in 2012 \citep[see][]{2015ApJ...815...55R,2018ApJ...854...49M} and  $N_\mathrm{H,los} \sim (3.6\pm0.4) \times 10^{23} \mathrm{cm}^{-2} $ in 2016 \citep{2018ApJ...854...42B} based on the \nustar\, data. Besides, NGC 7582 was type 1 in 1977 and experienced a fast optical type change from type 1 to type 1.8 in 1998, then to type 2 between 2004 and 2016 \citep[see][]{2022ApJS..261....4O,2024arXiv241108676J}. 

\section{Fitting Results}
\begin{table*}
\scriptsize
\centering
\caption{The parameters of five CLAGNs fitted with \texttt{pexrav} model. The table lists the Name, ObsID, cross constant of instruments (FPMB and soft X-ray observations) relative to FPMA,  line-of-sight column density in the unit of $10^{22}  \mathrm{\,cm}^{-2}$, photon index, cutoff energy, normalization in units of photons/keV/cm$^2$/s, line energy, sigma, and normalization the \fek\,, reflection factor (R), scatter fraction (fscat), $\chi^{2}$/$dof$, and the observed flux, and intrinsic flux at 2-10 keV band in units of erg/s/cm$^{-2}$. $^f$ means the value is fixed during the fitting.    \label{tab:slab}}
\begin{tabular}{lcccc ccccc  cccc ccl} \hline\hline
{Name} & {ObsID} & {Cross constant}& {$N_{\rm H,los}$ }      & {$\Gamma$} & {Ecut [keV]}  & {Norm} & {zgauss}  &{sigma} &
\\ {} & {}    &  {zgaussnorm}   & {R} &  {fscat}   & {$\chi^{2}_{r} $($\chi^{2}$/dof)} & {log$f_{\rm obs, 2-10keV} $} & {log$f_{\rm int, 2-10keV}$ }  \\ \hline
%
\multirow{2}*{ESO362-G18} & \multirow{2}*{60201046002/790810101} & 0.980$\pm$0.008/0.669$\pm$0.005 & 0.00$^f$ & 1.49 $\pm$ 0.00 & 200$^f$ & 2.11e-03 $\pm$ 2.14e-05 & 6.38 $\pm$ 0.01 & 0.05$\pm$0.01 & \\ 
 & &  2.91e-05 $\pm$ 1.21e-06 & 0.00$\pm$0.00& 0.00$\pm$0.00& 1.14(3104/2729) & -10.931 $\pm$ 0.001 & -10.942 $\pm$ 0.001 & \\
\multirow{2}*{NGC1365} & \multirow{2}*{60002046002/692840201} & 1.018$\pm$0.013/0.901$\pm$0.010 & 21.15$\pm$0.71 & 1.80 $\pm$ 0.03 & 200$^f$ & 6.42e-03 $\pm$ 4.01e-04 & 6.35 $\pm$ 0.01 & 0.05$^f$ & \\ 
 & &  2.10e-05 $\pm$ 1.16e-06 & -1.10$\pm$0.16& 0.00$\pm$0.01& 1.30(2684/2064) & -11.017 $\pm$ 0.003 & -10.790 $\pm$ 0.004 & \\
\multirow{2}*{NGC1365} & \multirow{2}*{60002046003/35458003} & 1.015$\pm$0.012/0.633$\pm$0.058 & 15.08$\pm$1.31 & 1.65 $\pm$ 0.09 & 289$\pm$211 & 4.30e-03 $\pm$ 8.08e-04 & 6.20 $\pm$ 0.09 & 0.19$\pm$0.07 & \\ 
 & &  2.70e-05 $\pm$ 5.95e-06 & -1.00$\pm$0.42& 0.05$\pm$0.01& 1.16(905/779) & -10.964 $\pm$ 0.002 & -10.726 $\pm$ 0.006 & \\
\multirow{2}*{NGC1365} & \multirow{2}*{60702058002/96123002} & 1.032$\pm$0.013/0.490$\pm$0.049 & 21.28$\pm$2.53 & 1.65 $\pm$ 0.05 & 200$^f$ & 2.49e-03 $\pm$ 2.03e-04 & 6.28 $\pm$ 0.03 & 0.05$^f$ & \\ 
 & &  1.89e-05 $\pm$ 2.07e-06 & -2.46$\pm$0.41& 0.05$\pm$0.03& 1.08(815/754) & -11.171 $\pm$ 0.003 & -10.979 $\pm$ 0.005 & \\
\multirow{2}*{NGC1365} & \multirow{2}*{60702058004/89213003} & 1.034$\pm$0.015/1.247$\pm$0.111 & 25.81$\pm$3.07 & 1.79 $\pm$ 0.05 & 200$^f$ & 3.49e-03 $\pm$ 3.30e-04 & 6.21 $\pm$ 0.06 & 0.05$^f$ & \\ 
 & &  1.74e-05 $\pm$ 2.60e-06 & -3.21$\pm$0.66& 0.04$\pm$0.03& 1.03(590/572) & -11.155 $\pm$ 0.003 & -10.924 $\pm$ 0.005 & \\
\multirow{2}*{NGC1365} & \multirow{2}*{60702058006} & 1.025$\pm$0.031& 36.61$\pm$10.50 & 1.76 $\pm$ 0.10 & 200$^f$ & 2.07e-03 $\pm$ 6.00e-04 & 6.34 $\pm$ 0.07 & 0.05$^f$ & \\ 
 & &  1.33e-05 $\pm$ 4.14e-06 & -5.69$\pm$3.07& 0.00$\pm$0.06& 1.01(164/163) & -11.331 $\pm$ 0.007 & -11.126 $\pm$ 0.006 & \\
\multirow{2}*{NGC1365} & \multirow{2}*{60702058008} & 1.037$\pm$0.014& 13.95$\pm$0.81 & 1.70 $\pm$ 0.05 & 200$^f$ & 3.11e-03 $\pm$ 2.69e-04 & 6.40 $\pm$ 0.03 & 0.05$^f$ & \\ 
 & &  2.03e-05 $\pm$ 2.49e-06 & -2.10$\pm$0.38& 0.00$\pm$3.86& 1.16(778/670) & -11.099 $\pm$ 0.003 & -10.906 $\pm$ 0.039 & \\
\multirow{2}*{NGC1365} & \multirow{2}*{60702058010} & 1.055$\pm$0.015& 28.84$\pm$5.16 & 1.53 $\pm$ 0.14 & 94$\pm$35 & 1.73e-03 $\pm$ 4.90e-04 & 6.28 $\pm$ 0.03 & 0.05$^f$ & \\ 
 & &  1.59e-05 $\pm$ 2.48e-06 & -3.34$\pm$0.73& 0.05$\pm$0.03& 1.07(627/586) & -11.252 $\pm$ 0.003 & -11.063 $\pm$ 0.070 & \\
\multirow{2}*{NGC1365} & \multirow{2}*{60702058012} & 1.004$\pm$0.017& 87.16$\pm$14.27 & 1.53 $\pm$ 0.16 & 149$\pm$97 & 1.85e-03 $\pm$ 9.62e-04 & 6.33 $\pm$ 0.02 & 0.05$^f$ & \\ 
 & &  1.52e-05 $\pm$ 1.75e-06 & -2.37$\pm$0.86& 0.05$\pm$0.03& 1.08(557/515) & -11.506 $\pm$ 0.004 & -11.026 $\pm$ 0.009 & \\
\multirow{2}*{NGC4151} & \multirow{2}*{60001111002/80073001} & 0.980$\pm$0.004/0.577$\pm$0.022 & 5.23$\pm$0.24 & 1.51 $\pm$ 0.02 & 64$\pm$3 & 4.89e-02 $\pm$ 2.26e-03 & 6.26 $\pm$ 0.02 & 0.20$\pm$0.03 & \\ 
 & &  3.13e-04 $\pm$ 2.83e-05 & -0.67$\pm$0.06& 0.03$\pm$0.01& 1.11(1715/1550) & -9.676 $\pm$ 0.001 & -9.677 $\pm$ 0.007 & \\
\multirow{2}*{NGC4151} & \multirow{2}*{60001111003/80073001} & 0.997$\pm$0.002/0.768$\pm$0.028 & 7.07$\pm$0.20 & 1.39 $\pm$ 0.02 & 62$\pm$2 & 3.73e-02 $\pm$ 1.29e-03 & 6.26 $\pm$ 0.01 & 0.22$\pm$0.01 & \\ 
 & &  4.10e-04 $\pm$ 1.80e-05 & -0.47$\pm$0.03& 0.04$\pm$0.01& 1.12(2287/2035) & -9.743 $\pm$ 0.001 & -9.611 $\pm$ 0.003 & \\
\multirow{2}*{NGC4151} & \multirow{2}*{60001111005} & 0.999$\pm$0.002& 17.67$\pm$0.75 & 1.70 $\pm$ 0.02 & 126$\pm$10 & 6.45e-02 $\pm$ 2.97e-03 & 6.31 $\pm$ 0.01 & 0.05$^f$ & \\ 
 & &  2.26e-04 $\pm$ 1.10e-05 & -0.78$\pm$0.04& 0.39$\pm$0.01& 1.16(2333/2020) & -9.663 $\pm$ 0.000 & -9.601 $\pm$ 0.003 & \\
\multirow{2}*{NGC4151} & \multirow{2}*{60502017002} & 0.988$\pm$0.003& 17.86$\pm$0.99 & 1.70 $\pm$ 0.03 & 141$\pm$18 & 6.37e-02 $\pm$ 4.11e-03 & 6.27 $\pm$ 0.01 & 0.05$^f$ & \\ 
 & &  2.38e-04 $\pm$ 1.43e-05 & -0.52$\pm$0.05& 0.27$\pm$0.01& 1.07(1723/1613) & -9.752 $\pm$ 0.001 & -9.603 $\pm$ 0.004 & \\
\multirow{2}*{NGC4151} & \multirow{2}*{60502017004/88889001} & 0.999$\pm$0.002& 14.30$\pm$0.92 & 1.76 $\pm$ 0.02 & 187$\pm$24 & 7.40e-02 $\pm$ 3.49e-03 & 6.29 $\pm$ 0.01 & 0.05$^f$ & \\ 
 & &  2.39e-04 $\pm$ 1.39e-05 & -0.72$\pm$0.04& 0.44$\pm$0.02& 1.02(1890/1847) & -9.595 $\pm$ 0.001 & -9.574 $\pm$ 0.004 & \\
\multirow{2}*{NGC4151} & \multirow{2}*{60502017006} & 0.992$\pm$0.003& 13.61$\pm$1.25 & 1.75 $\pm$ 0.02 & 150$\pm$19 & 6.24e-02 $\pm$ 3.71e-03 & 6.29 $\pm$ 0.01 & 0.05$^f$ & \\ 
 & &  2.53e-04 $\pm$ 1.60e-05 & -0.81$\pm$0.06& 0.58$\pm$0.03& 1.03(1684/1637) & -9.599 $\pm$ 0.001 & -9.648 $\pm$ 0.006 & \\
\multirow{2}*{NGC4151} & \multirow{2}*{60502017008/88889004} & 0.984$\pm$0.003& 18.91$\pm$1.14 & 1.71 $\pm$ 0.03 & 130$\pm$17 & 5.73e-02 $\pm$ 4.23e-03 & 6.29 $\pm$ 0.01 & 0.05$^f$ & \\ 
 & &  2.33e-04 $\pm$ 1.46e-05 & -0.77$\pm$0.06& 0.34$\pm$0.02& 1.11(1727/1551) & -9.755 $\pm$ 0.001 & -9.657 $\pm$ 0.005 & \\
\multirow{2}*{NGC4151} & \multirow{2}*{60502017010} & 0.976$\pm$0.003& 19.67$\pm$1.16 & 1.73 $\pm$ 0.03 & 168$\pm$28 & 6.13e-02 $\pm$ 4.70e-03 & 6.29 $\pm$ 0.01 & 0.05$^f$ & \\ 
 & &  2.23e-04 $\pm$ 1.52e-05 & -0.62$\pm$0.06& 0.39$\pm$0.02& 1.09(1688/1549) & -9.725 $\pm$ 0.001 & -9.638 $\pm$ 0.005 & \\
\multirow{2}*{NGC4151} & \multirow{2}*{60502017012/88889005} & 0.978$\pm$0.004& 17.81$\pm$1.31 & 1.70 $\pm$ 0.03 & 201$\pm$41 & 5.25e-02 $\pm$ 4.21e-03 & 6.31 $\pm$ 0.01 & 0.05$^f$ & \\ 
 & &  2.81e-04 $\pm$ 1.59e-05 & -0.59$\pm$0.06& 0.47$\pm$0.02& 1.04(1592/1527) & -9.717 $\pm$ 0.001 & -9.686 $\pm$ 0.006 & \\
\multirow{2}*{NGC5548} & \multirow{2}*{60002044002/91744028} & 1.002$\pm$0.008/0.750$\pm$0.042 & 1.37$\pm$0.44 & 1.57 $\pm$ 0.03 & 196$\pm$79 & 9.33e-03 $\pm$ 7.19e-04 & 6.20 $\pm$ 0.19 & 0.45$\pm$0.11 & \\ 
 & &  7.41e-05 $\pm$ 1.67e-05 & -0.10$\pm$0.07& 0.05$\pm$0.11& 1.01(896/890) & -10.361 $\pm$ 0.002 & -10.340 $\pm$ 0.043 & \\
\multirow{2}*{NGC5548} & \multirow{2}*{60002044003/91711077} & 1.015$\pm$0.008/0.795$\pm$0.051 & 1.76$\pm$0.35 & 1.69 $\pm$ 0.03 & 165$\pm$50 & 1.10e-02 $\pm$ 6.01e-04 & 6.35 $\pm$ 1.28 & 0.00$\pm$0.03 & \\ 
 & &  3.20e-05 $\pm$ 5.88e-06 & -0.38$\pm$0.09& 0.05$\pm$0.05& 1.04(902/864) & -10.383 $\pm$ 0.002 & -10.361 $\pm$ 0.036 & \\
\multirow{2}*{NGC5548} & \multirow{2}*{60002044005/91744040} & 0.985$\pm$0.006/0.094$\pm$0.049 & 2.86$\pm$1.10 & 1.55 $\pm$ 0.04 & 113$\pm$21 & 7.77e-03 $\pm$ 2.02e-03 & 6.32 $\pm$ 0.02 & 0.06$\pm$0.06 & \\ 
 & &  4.68e-05 $\pm$ 5.56e-06 & -0.25$\pm$0.09& 0.05$\pm$0.29& 0.96(1044/1086) & -10.471 $\pm$ 0.001 & -10.433 $\pm$ 0.025 & \\
\multirow{2}*{NGC5548} & \multirow{2}*{60002044006/91711139} & 1.017$\pm$0.006/0.094$\pm$0.044 & 0.89$\pm$1.21 & 1.64 $\pm$ 0.04 & 119$\pm$23 & 9.74e-03 $\pm$ 1.06e-02 & 6.35 $\pm$ 0.03 & 0.19$\pm$0.04 & \\ 
 & &  5.57e-05 $\pm$ 7.39e-06 & -0.35$\pm$0.40& 0.05$\pm$1.17& 1.03(1121/1092) & -10.375 $\pm$ 0.001 & -10.404 $\pm$ 0.055 & \\
\multirow{2}*{NGC5548} & \multirow{2}*{60002044008/80131001} & 0.977$\pm$0.007/0.692$\pm$0.044 & 1.88$\pm$0.42 & 1.40 $\pm$ 0.03 & 64$\pm$6 & 5.15e-03 $\pm$ 3.26e-04 & 6.35 $\pm$ 0.02 & 0.14$\pm$0.04 & \\ 
 & &  5.56e-05 $\pm$ 5.64e-06 & -0.29$\pm$0.07& 0.05$\pm$0.07& 1.04(1132/1084) & -10.512 $\pm$ 0.001 & -10.488 $\pm$ 0.029 & \\
\multirow{2}*{NGC5548} & \multirow{2}*{90701601002/95671036} & 1.007$\pm$0.008/0.819$\pm$0.059 & 0.85$\pm$0.48 & 1.63 $\pm$ 0.03 & 175$\pm$54 & 7.95e-03 $\pm$ 1.79e-03 & 6.34 $\pm$ 0.03 & 0.14$\pm$0.04 & \\ 
 & &  5.35e-05 $\pm$ 6.48e-06 & -0.33$\pm$0.13& 0.05$\pm$0.28& 0.98(940/959) & -10.447 $\pm$ 0.002 & -10.433 $\pm$ 0.038 & \\
\multirow{2}*{NGC7582} & \multirow{2}*{60061318002/32534001} & 0.968$\pm$0.021/0.960$\pm$0.106 & 24.89$\pm$2.18 & 1.32 $\pm$ 0.09 & 200$^f$ & 1.68e-03 $\pm$ 3.04e-04 & 6.36 $\pm$ 0.03 & 0.05$^f$ & \\ 
 & &  2.64e-05 $\pm$ 3.94e-06 & -1.00$\pm$0.78& 0.05$\pm$0.01& 1.12(336/301) & -11.177 $\pm$ 0.005 & -11.110 $\pm$ 0.039 & \\
\multirow{2}*{NGC7582} & \multirow{2}*{60061318004} & 0.987$\pm$0.026& 51.18$\pm$13.15 & 1.42 $\pm$ 0.08 & 104$^f$ & 1.08e-03 $\pm$ 3.83e-04 & 6.37 $\pm$ 0.03 & 0.05$^f$ & \\ 
 & &  2.32e-05 $\pm$ 3.34e-06 & -4.96$\pm$2.76& 0.10$\pm$0.03& 1.09(231/212) & -11.349 $\pm$ 0.006 & -11.192 $\pm$ 0.070 & \\
\multirow{2}*{NGC7582} & \multirow{2}*{60201003002/782720301} & 0.992$\pm$0.009/0.676$\pm$0.008 & 34.51$\pm$0.79 & 1.62 $\pm$ 0.03 & 200$^f$ & 9.63e-03 $\pm$ 7.30e-04 & 6.39 $\pm$ 0.01 & 0.05$^f$ & \\ 
 & &  2.95e-05 $\pm$ 1.98e-06 & -0.29$\pm$0.09& 0.02$\pm$0.00& 1.41(3880/2754) & -10.912 $\pm$ 0.002 & -10.546 $\pm$ 0.009 & \\
 \hline
 \end{tabular}
\end{table*}

\begin{table*}
\scriptsize
\caption{The parameters fitted with the xclumpy model. The table lists the Name, ObsID, cross constant of instruments (FPMB and soft X-ray observations relative to FPMA,  line-of-sight column density, photon index, scatter fraction (fscat), inclination angle, $\chi^{2}$/$dof$, and the intrinsic 2-10 keV flux and luminosity.  \label{tab:xclumpy}}
\begin{tabular}{lcccc ccccc  ccccl}\hline\hline
{Name} & {ObsID} & {Cross constant} & {$N_{\rm H,los}$ [$10^{22}  \mathrm{\,cm}^{-2}$]}  & {$  N_{\rm H,tor}$  [$ 10^{24} \mathrm{\,cm}^{-2}$]}      & {$\Gamma$}    & {Norm [photons/keV/cm$^2$/s] } & {Inclination angle} \\  
{}  & {} & {sigma degree} & {AL}  &  {fscat}   & {$\chi^{2}_{r} $($\chi^{2}$/dof)} & {log$f_{\rm int, 2-10keV} $} & {log$L_{\rm int, 2-10keV}$} \\ \hline  
                      

\multirow{2}*{ESO362-G18} & \multirow{2}*{60201046002} & 0.981$\pm$0.008& 1.2$\pm$0.4& 3.8 $\pm$ 1.3 & 1.68 $\pm$ 0.04 & 2.91e-03 $\pm$ 2.22e-04 & 45 $\pm$ 10 & \\ 
 & & 18.5 $\pm$ 4.6 & 1.7 $\pm$ 0.6 & 0.00$\pm$-1.00& 0.87(417/479) & -10.921 $\pm$ 0.017 & 42.616 $\pm$ 0.017 \\ \hline
\multirow{2}*{NGC1365} & \multirow{2}*{60002046002} & 1.020$\pm$0.015& 19.2$\pm$7.1& 10.0 $\pm$ 3.7 & 1.66 $\pm$ 0.04 & 4.39e-03 $\pm$ 5.09e-04 & 50 $\pm$ 2 & \\ 
 & & 19.8 $\pm$ 1.7 & 0.9 $\pm$ 0.3 & 0.09$\pm$0.03& 1.10(380/346) & -10.724 $\pm$ 0.076 & 42.102 $\pm$ 0.076 \\
\multirow{2}*{NGC1365} & \multirow{2}*{60002046003} & 1.010$\pm$0.012& 15.2$\pm$5.6& 10.0 $\pm$ 3.7 & 1.64 $\pm$ 0.04 & 4.77e-03 $\pm$ 4.15e-04 & 50 $\pm$ 2 & \\ 
 & & 19.2 $\pm$ 1.7 & 0.9 $\pm$ 0.3 & 0.04$\pm$0.05& 1.15(448/391) & -10.674 $\pm$ 0.048 & 42.152 $\pm$ 0.048 \\
\multirow{2}*{NGC1365} & \multirow{2}*{60702058002} & 1.030$\pm$0.015& 28.6$\pm$10.6& 10.0 $\pm$ 3.7 & 1.49 $\pm$ 0.04 & 2.48e-03 $\pm$ 3.34e-04 & 50 $\pm$ 2 & \\ 
 & & 20.9 $\pm$ 1.9 & 2.0 $\pm$ 0.3 & 0.14$\pm$0.02& 1.18(458/387) & -10.854 $\pm$ 0.123 & 41.972 $\pm$ 0.123 \\
\multirow{2}*{NGC1365} & \multirow{2}*{60702058004} & 1.040$\pm$0.018& 34.3$\pm$12.7& 10.0 $\pm$ 3.7 & 1.62 $\pm$ 0.06 & 4.05e-03 $\pm$ 7.46e-04 & 50 $\pm$ 2 & \\ 
 & & 21.4 $\pm$ 2.0 & 1.1 $\pm$ 0.3 & 0.10$\pm$0.01& 1.14(333/293) & -10.731 $\pm$ 0.138 & 42.095 $\pm$ 0.138 \\
\multirow{2}*{NGC1365} & \multirow{2}*{60702058006} & 1.020$\pm$0.036& 55.9$\pm$20.7& 10.0 $\pm$ 3.7 & 1.60 $\pm$ 0.09 & 4.31e-03 $\pm$ 1.38e-03 & 50 $\pm$ 2 & \\ 
 & & 23.2 $\pm$ 2.2 & 1.2 $\pm$ 0.4 & 0.07$\pm$0.02& 1.03(87/85) & -10.690 $\pm$ 0.330 & 42.136 $\pm$ 0.330 \\
\multirow{2}*{NGC1365} & \multirow{2}*{60702058008} & 1.040$\pm$0.014& 12.9$\pm$4.8& 10.0 $\pm$ 3.7 & 1.48 $\pm$ 0.03 & 2.46e-03 $\pm$ 1.94e-04 & 50 $\pm$ 2 & \\ 
 & & 18.9 $\pm$ 1.7 & 1.7 $\pm$ 0.5 & 0.00$\pm$0.03& 1.20(416/347) & -10.850 $\pm$ 0.060 & 41.976 $\pm$ 0.060 \\
\multirow{2}*{NGC1365} & \multirow{2}*{60702058010} & 1.060$\pm$0.018& 51.0$\pm$18.9& 10.0 $\pm$ 3.7 & 1.63 $\pm$ 0.06 & 4.66e-03 $\pm$ 8.99e-04 & 50 $\pm$ 2 & \\ 
 & & 22.8 $\pm$ 2.1 & 1.1 $\pm$ 0.2 & 0.08$\pm$0.01& 1.14(344/302) & -10.677 $\pm$ 0.153 & 42.149 $\pm$ 0.153 \\
\multirow{2}*{NGC1365} & \multirow{2}*{60702058012} & 0.998$\pm$0.020& 91.5$\pm$33.9& 10.0 $\pm$ 3.7 & 1.63 $\pm$ 0.06 & 4.91e-03 $\pm$ 1.07e-03 & 50 $\pm$ 2 & \\ 
 & & 25.5 $\pm$ 2.6 & 1.3 $\pm$ 0.2 & 0.06$\pm$0.01& 1.01(270/269) & -10.654 $\pm$ 0.159 & 42.172 $\pm$ 0.159 \\ \hline
\multirow{2}*{NGC4151} & \multirow{2}*{60001111002} & 0.960$\pm$0.006& 23.7$\pm$1.9& 4.5 $\pm$ 0.4 & 1.84 $\pm$ 0.01 & 8.26e-02 $\pm$ 4.49e-03 & 48 $\pm$ 1 & \\ 
 & & 24.2 $\pm$ 1.2 & 0.5 $\pm$ 0.1 & 0.44$\pm$0.02& 1.08(806/748) & -9.569 $\pm$ 0.042 & 42.812 $\pm$ 0.042 \\
\multirow{2}*{NGC4151} & \multirow{2}*{60001111003} & 0.995$\pm$0.003& 21.4$\pm$1.7& 4.5 $\pm$ 0.4 & 1.78 $\pm$ 0.01 & 7.59e-02 $\pm$ 2.24e-03 & 48 $\pm$ 1 & \\ 
 & & 23.8 $\pm$ 1.2 & 0.8 $\pm$ 0.0 & 0.28$\pm$0.01& 1.03(1015/987) & -9.566 $\pm$ 0.022 & 42.815 $\pm$ 0.022 \\
\multirow{2}*{NGC4151} & \multirow{2}*{60001111005} & 0.998$\pm$0.003& 19.1$\pm$1.5& 4.5 $\pm$ 0.4 & 1.80 $\pm$ 0.01 & 8.19e-02 $\pm$ 2.29e-03 & 48 $\pm$ 1 & \\ 
 & & 23.3 $\pm$ 1.2 & 0.7 $\pm$ 0.0 & 0.35$\pm$0.01& 1.15(1168/1018) & -9.546 $\pm$ 0.020 & 42.835 $\pm$ 0.020 \\
\multirow{2}*{NGC4151} & \multirow{2}*{60502017002} & 0.981$\pm$0.004& 18.6$\pm$1.5& 4.5 $\pm$ 0.4 & 1.83 $\pm$ 0.01 & 8.54e-02 $\pm$ 3.01e-03 & 48 $\pm$ 1 & \\ 
 & & 23.2 $\pm$ 1.2 & 0.7 $\pm$ 0.1 & 0.23$\pm$0.01& 1.09(892/819) & -9.548 $\pm$ 0.028 & 42.833 $\pm$ 0.028 \\
\multirow{2}*{NGC4151} & \multirow{2}*{60502017004} & 0.997$\pm$0.004& 13.2$\pm$1.1& 4.5 $\pm$ 0.4 & 1.79 $\pm$ 0.01 & 8.15e-02 $\pm$ 1.90e-03 & 48 $\pm$ 1 & \\ 
 & & 22.1 $\pm$ 1.1 & 0.8 $\pm$ 0.1 & 0.38$\pm$0.02& 1.03(964/938) & -9.541 $\pm$ 0.018 & 42.839 $\pm$ 0.018 \\
\multirow{2}*{NGC4151} & \multirow{2}*{60502017006} & 0.985$\pm$0.005& 14.1$\pm$1.1& 4.5 $\pm$ 0.4 & 1.81 $\pm$ 0.01 & 7.41e-02 $\pm$ 2.45e-03 & 48 $\pm$ 1 & \\ 
 & & 22.3 $\pm$ 1.1 & 0.9 $\pm$ 0.1 & 0.51$\pm$0.03& 1.09(911/833) & -9.596 $\pm$ 0.027 & 42.785 $\pm$ 0.027 \\
\multirow{2}*{NGC4151} & \multirow{2}*{60502017008} & 0.978$\pm$0.005& 19.8$\pm$1.6& 4.5 $\pm$ 0.4 & 1.80 $\pm$ 0.01 & 7.12e-02 $\pm$ 2.79e-03 & 48 $\pm$ 1 & \\ 
 & & 23.4 $\pm$ 1.2 & 0.8 $\pm$ 0.1 & 0.31$\pm$0.01& 1.07(843/785) & -9.607 $\pm$ 0.033 & 42.774 $\pm$ 0.033 \\
\multirow{2}*{NGC4151} & \multirow{2}*{60502017010} & 0.961$\pm$0.005& 19.5$\pm$1.6& 4.5 $\pm$ 0.4 & 1.81 $\pm$ 0.01 & 7.39e-02 $\pm$ 3.08e-03 & 48 $\pm$ 1 & \\ 
 & & 23.4 $\pm$ 1.2 & 0.8 $\pm$ 0.1 & 0.34$\pm$0.01& 1.12(883/788) & -9.597 $\pm$ 0.034 & 42.783 $\pm$ 0.034 \\
\multirow{2}*{NGC4151} & \multirow{2}*{60502017012} & 0.964$\pm$0.006& 16.7$\pm$1.3& 4.5 $\pm$ 0.4 & 1.76 $\pm$ 0.01 & 5.97e-02 $\pm$ 2.34e-03 & 48 $\pm$ 1 & \\ 
 & & 22.8 $\pm$ 1.1 & 1.2 $\pm$ 0.1 & 0.41$\pm$0.02& 1.02(798/780) & -9.656 $\pm$ 0.034 & 42.724 $\pm$ 0.034 \\ \hline
\multirow{2}*{NGC5548} & \multirow{2}*{60002044002} & 1.000$\pm$0.024& 18.1$\pm$2.7& 4.0 $\pm$ 0.6 & 1.80 $\pm$ 0.04 & 7.38e-03 $\pm$ 1.40e-03 & 30 $\pm$ 8 & \\ 
 & & 33.9 $\pm$ 5.2 & 0.7 $\pm$ 0.3 & 1.39$\pm$0.21& 0.98(417/428) & -10.602 $\pm$ 0.187 & 43.221 $\pm$ 0.187 \\
\multirow{2}*{NGC5548} & \multirow{2}*{60002044003} & 1.020$\pm$0.012& 3.7$\pm$0.6& 4.0 $\pm$ 0.6 & 1.81 $\pm$ 0.04 & 1.09e-02 $\pm$ 3.33e-03 & 30 $\pm$ 8 & \\ 
 & & 27.5 $\pm$ 4.6 & 0.5 $\pm$ 0.2 & 0.33$\pm$0.34& 0.99(409/416) & -10.439 $\pm$ 0.137 & 43.384 $\pm$ 0.137 \\
\multirow{2}*{NGC5548} & \multirow{2}*{60002044005} & 0.958$\pm$0.014& 16.3$\pm$2.5& 4.0 $\pm$ 0.6 & 1.78 $\pm$ 0.03 & 7.14e-03 $\pm$ 7.88e-04 & 30 $\pm$ 8 & \\ 
 & & 33.3 $\pm$ 5.1 & 1.1 $\pm$ 0.2 & 0.94$\pm$0.09& 0.94(505/540) & -10.603 $\pm$ 0.132 & 43.220 $\pm$ 0.132 \\
\multirow{2}*{NGC5548} & \multirow{2}*{60002044006} & 1.020$\pm$0.008& 2.1$\pm$0.3& 4.0 $\pm$ 0.6 & 1.81 $\pm$ 0.03 & 1.20e-02 $\pm$ 2.75e-03 & 30 $\pm$ 8 & \\ 
 & & 26.1 $\pm$ 4.2 & 0.8 $\pm$ 0.2 & 0.16$\pm$0.22& 1.05(564/539) & -10.398 $\pm$ 0.097 & 43.425 $\pm$ 0.097 \\
\multirow{2}*{NGC5548} & \multirow{2}*{60002044008} & 0.948$\pm$0.014& 21.2$\pm$3.2& 4.0 $\pm$ 0.6 & 1.80 $\pm$ 0.03 & 7.97e-03 $\pm$ 9.91e-04 & 30 $\pm$ 8 & \\ 
 & & 34.7 $\pm$ 5.3 & 0.9 $\pm$ 0.2 & 0.79$\pm$0.07& 0.91(488/535) & -10.568 $\pm$ 0.132 & 43.254 $\pm$ 0.132 \\
\multirow{2}*{NGC5548} & \multirow{2}*{90701601002} & 1.010$\pm$0.009& 2.3$\pm$0.4& 4.0 $\pm$ 0.6 & 1.78 $\pm$ 0.03 & 9.34e-03 $\pm$ 2.65e-03 & 30 $\pm$ 8 & \\ 
 & & 26.3 $\pm$ 4.3 & 1.0 $\pm$ 0.2 & 0.20$\pm$0.28& 0.93(440/474) & -10.486 $\pm$ 0.122 & 43.337 $\pm$ 0.122 \\ \hline
\multirow{2}*{NGC7582} & \multirow{2}*{60061318002} & 0.956$\pm$0.027& 43.1$\pm$33.1& 9.9 $\pm$ 7.7 & 1.39 $\pm$ 0.07 & 2.47e-03 $\pm$ 8.78e-04 & 60 $\pm$ 10 & \\ 
 & & 16.9 $\pm$ 4.0 & 3.6 $\pm$ 1.6 & 0.14$\pm$0.05& 1.12(169/151) & -10.785 $\pm$ 0.150 & 42.009 $\pm$ 0.150 \\
\multirow{2}*{NGC7582} & \multirow{2}*{60061318004} & 0.985$\pm$0.032& 73.0$\pm$56.1& 9.9 $\pm$ 7.7 & 1.44 $\pm$ 0.10 & 3.28e-03 $\pm$ 1.60e-03 & 60 $\pm$ 10 & \\ 
 & & 18.6 $\pm$ 4.7 & 2.7 $\pm$ 1.1 & 0.08$\pm$0.03& 1.14(128/113) & -10.697 $\pm$ 0.176 & 42.096 $\pm$ 0.176 \\
\multirow{2}*{NGC7582} & \multirow{2}*{60201003002} & 0.991$\pm$0.010& 32.0$\pm$24.6& 9.9 $\pm$ 7.7 & 1.64 $\pm$ 0.05 & 8.24e-03 $\pm$ 2.17e-03 & 60 $\pm$ 10 & \\ 
 & & 16.2 $\pm$ 3.8 & 1.2 $\pm$ 0.5 & 0.06$\pm$0.03& 1.06(519/491) & -10.436 $\pm$ 0.051 & 42.357 $\pm$ 0.051 \\
 \hline 
 \end{tabular}
\end{table*}

\begin{table*}
\scriptsize
\caption{The parameters fitted with the uxclumpy model. The table lists the Name, ObsID, cross constant of instruments (FPMB and soft X-ray observations relative to FPMA,  line-of-sight column density, photon index, scatter fraction (fscat), inclination angle, $\chi^{2}$/$dof$, and the intrinsic 2-10 keV flux and luminosity. $^f$ means the value is fixed during the fitting. $u$ means the scatter factor reaches the upper limit.  \label{tab:uxclumpy}}
\begin{tabular}{lcccc ccccc  cccccl}
{Name} & {ObsID} & {Cross constant} & {$N_{\rm H,los}$ [$10^{22}  \mathrm{\,cm}^{-2}$]}       & {$\Gamma$}   & {Norm [photons/keV/cm$^2$/s] }   & {Inclination angle}   & {torsigma}  & \\  
 {}  & {} & {CTKcover} & {fscat}   & {$\chi^{2}_{r} $($\chi^{2}$/dof)} & {log$f_{\rm int, 2-10keV} $} & {log$L_{\rm int, 2-10keV}$} \\ \hline 
                      

\multirow{2}*{ESO362-G18} & \multirow{2}*{60201046002} & 0.980$\pm$0.008& 0.02$\pm$0.00& 1.56 $\pm$ 0.00 & 2.43e-03 $\pm$ 5.70e-04 & 73 $\pm$ 6 & 13 $\pm$ 1 & \\ 
 & & 0.25 $\pm$ 0.02 & 0.00$^f$& 1.13(3094/2737) & -10.916 $\pm$ 0.102 & 42.621 $\pm$ 0.102 \\ \hline 
\multirow{2}*{NGC1365} & \multirow{2}*{60002046002} & 1.020$\pm$0.013& 12.45$\pm$1.50& 1.61 $\pm$ 0.05 & 4.64e-03 $\pm$ 3.41e-04 & 70 $\pm$ 9 & 84 $\pm$ -1 & \\ 
 & & 0.28 $\pm$ 0.01 & 0.05$^u$& 1.11(384/346) & -10.665 $\pm$ 0.047 & 42.161 $\pm$ 0.047 \\
\multirow{2}*{NGC1365} & \multirow{2}*{60002046003} & 1.010$\pm$0.011& 12.08$\pm$0.77& 1.53 $\pm$ 0.01 & 3.63e-03 $\pm$ 2.07e-04 & 70 $\pm$ 9 & 2 $\pm$ 1 & \\ 
 & & 0.28 $\pm$ 0.01 & 0.05$^u$& 1.21(473/391) & -10.716 $\pm$ 0.026 & 42.110 $\pm$ 0.026 \\
\multirow{2}*{NGC1365} & \multirow{2}*{60702058002} & 1.030$\pm$0.013& 15.06$\pm$1.59& 1.45 $\pm$ 0.05 & 2.51e-03 $\pm$ 2.11e-04 & 70 $\pm$ 9 & 84 $\pm$ 68 & \\ 
 & & 0.28 $\pm$ 0.01 & 0.05$^u$& 1.23(474/387) & -10.821 $\pm$ 0.051 & 42.005 $\pm$ 0.051 \\
\multirow{2}*{NGC1365} & \multirow{2}*{60702058004} & 1.030$\pm$0.015& 17.12$\pm$1.95& 1.50 $\pm$ 0.04 & 3.07e-03 $\pm$ 2.37e-04 & 70 $\pm$ 9 & 84 $\pm$ 74 & \\ 
 & & 0.28 $\pm$ 0.01 & 0.05$^u$& 1.19(350/293) & -10.768 $\pm$ 0.044 & 42.058 $\pm$ 0.044 \\
\multirow{2}*{NGC1365} & \multirow{2}*{60702058006} & 1.020$\pm$0.031& 25.72$\pm$4.49& 1.37 $\pm$ 0.05 & 2.16e-03 $\pm$ 2.81e-04 & 70 $\pm$ 9 & 84 $\pm$ -1 & \\ 
 & & 0.28 $\pm$ 0.01 & 0.05$^u$& 1.17(99/85) & -10.829 $\pm$ 0.067 & 41.997 $\pm$ 0.067 \\
\multirow{2}*{NGC1365} & \multirow{2}*{60702058008} & 1.040$\pm$0.014& 11.80$\pm$1.36& 1.54 $\pm$ 0.04 & 3.20e-03 $\pm$ 2.26e-04 & 70 $\pm$ 9 & 84 $\pm$ -1 & \\ 
 & & 0.28 $\pm$ 0.01 & 0.05$^u$& 1.21(420/347) & -10.778 $\pm$ 0.041 & 42.048 $\pm$ 0.041 \\
\multirow{2}*{NGC1365} & \multirow{2}*{60702058010} & 1.050$\pm$0.015& 23.13$\pm$2.03& 1.40 $\pm$ 0.03 & 2.40e-03 $\pm$ 1.67e-04 & 70 $\pm$ 9 & 84 $\pm$ 56 & \\ 
 & & 0.28 $\pm$ 0.01 & 0.05$^u$& 1.28(387/302) & -10.805 $\pm$ 0.037 & 42.021 $\pm$ 0.037 \\
\multirow{2}*{NGC1365} & \multirow{2}*{60702058012} & 1.000$\pm$0.017& 99.83$\pm$4.02& 1.60 $\pm$ 0.04 & 7.49e-03 $\pm$ 1.36e-03 & 70 $\pm$ 9 & 0 $\pm$ 5 & \\ 
 & & 0.28 $\pm$ 0.01 & 0.05$\pm$0.01& 1.17(314/269) & -10.450 $\pm$ 0.084 & 42.376 $\pm$ 0.084 \\ \hline 
\multirow{2}*{NGC4151} & \multirow{2}*{60001111002} & 0.979$\pm$0.004& 6.19$\pm$0.72& 1.72 $\pm$ 0.01 & 8.89e-02 $\pm$ 6.16e-03 & 78 $\pm$ 1 & 21 $\pm$ 3 & \\ 
 & & 0.30 $\pm$ 0.00 & 0.05$^u$& 1.35(1011/748) & -9.456 $\pm$ 0.031 & 42.924 $\pm$ 0.031 \\
\multirow{2}*{NGC4151} & \multirow{2}*{60001111003} & 0.997$\pm$0.002& 7.76$\pm$0.06& 1.60 $\pm$ 0.00 & 7.56e-02 $\pm$ 1.25e-03 & 78 $\pm$ 1 & 4 $\pm$ 0 & \\ 
 & & 0.30 $\pm$ 0.00 & 0.03$\pm$0.01& 1.63(1610/987) & -9.445 $\pm$ 0.007 & 42.936 $\pm$ 0.007 \\
\multirow{2}*{NGC4151} & \multirow{2}*{60001111005} & 1.000$\pm$0.002& 7.38$\pm$0.28& 1.72 $\pm$ 0.01 & 9.79e-02 $\pm$ 1.41e-03 & 78 $\pm$ 1 & 25 $\pm$ 1 & \\ 
 & & 0.30 $\pm$ 0.00 & 0.05$^u$& 1.73(1761/1018) & -9.414 $\pm$ 0.009 & 42.966 $\pm$ 0.009 \\
\multirow{2}*{NGC4151} & \multirow{2}*{60502017002} & 0.988$\pm$0.003& 9.19$\pm$0.47& 1.75 $\pm$ 0.01 & 9.68e-02 $\pm$ 2.25e-03 & 78 $\pm$ 1 & 27 $\pm$ 1 & \\ 
 & & 0.30 $\pm$ 0.00 & 0.05$^u$& 1.39(1135/819) & -9.440 $\pm$ 0.012 & 42.941 $\pm$ 0.012 \\
\multirow{2}*{NGC4151} & \multirow{2}*{60502017004} & 0.999$\pm$0.002& 6.05$\pm$0.42& 1.76 $\pm$ 0.01 & 1.19e-01 $\pm$ 5.48e-03 & 78 $\pm$ 1 & 22 $\pm$ 2 & \\ 
 & & 0.30 $\pm$ 0.00 & 0.05$^u$& 1.27(1192/938) & -9.357 $\pm$ 0.021 & 43.024 $\pm$ 0.021 \\
\multirow{2}*{NGC4151} & \multirow{2}*{60502017006} & 0.995$\pm$0.003& 4.49$\pm$0.18& 1.69 $\pm$ 0.01 & 9.19e-02 $\pm$ 2.52e-03 & 78 $\pm$ 1 & 7 $\pm$ 0 & \\ 
 & & 0.30 $\pm$ 0.00 & 0.05$\pm$0.05& 1.44(1198/833) & -9.422 $\pm$ 0.014 & 42.959 $\pm$ 0.014 \\
\multirow{2}*{NGC4151} & \multirow{2}*{60502017008} & 0.985$\pm$0.003& 8.24$\pm$0.50& 1.74 $\pm$ 0.01 & 8.62e-02 $\pm$ 2.49e-03 & 78 $\pm$ 1 & 38 $\pm$ 10 & \\ 
 & & 0.30 $\pm$ 0.00 & 0.05$^u$& 1.41(1103/785) & -9.483 $\pm$ 0.014 & 42.897 $\pm$ 0.014 \\
\multirow{2}*{NGC4151} & \multirow{2}*{60502017010} & 0.976$\pm$0.003& 7.52$\pm$0.45& 1.73 $\pm$ 0.01 & 8.68e-02 $\pm$ 2.03e-03 & 78 $\pm$ 1 & 24 $\pm$ 2 & \\ 
 & & 0.30 $\pm$ 0.00 & 0.05$^u$& 1.43(1127/788) & -9.474 $\pm$ 0.012 & 42.907 $\pm$ 0.012 \\
\multirow{2}*{NGC4151} & \multirow{2}*{60502017012} & 0.978$\pm$0.004& 6.70$\pm$0.49& 1.75 $\pm$ 0.01 & 8.96e-02 $\pm$ 5.68e-03 & 78 $\pm$ 1 & 46 $\pm$ 16 & \\ 
 & & 0.30 $\pm$ 0.00 & 0.05$^u$& 1.28(997/780) & -9.473 $\pm$ 0.028 & 42.907 $\pm$ 0.028 \\ \hline 
\multirow{2}*{NGC5548} & \multirow{2}*{60002044002} & 1.000$\pm$0.008& 2.51$\pm$0.17& 1.76 $\pm$ 0.01 & 1.84e-02 $\pm$ 5.41e-03 & 38 $\pm$ 26 & 19 $\pm$ 5 & \\ 
 & & 0.00 $\pm$ -1.00 & 0.05$^u$& 1.00(427/428) & -10.178 $\pm$ 0.128 & 43.645 $\pm$ 0.128 \\
\multirow{2}*{NGC5548} & \multirow{2}*{60002044003} & 1.010$\pm$0.008& 1.93$\pm$0.24& 1.78 $\pm$ 0.02 & 1.55e-02 $\pm$ 2.42e-03 & 38 $\pm$ 26 & 26 $\pm$ 1 & \\ 
 & & 0.00 $\pm$ -1.00 & 0.05$^u$& 1.00(415/416) & -10.266 $\pm$ 0.069 & 43.557 $\pm$ 0.069 \\
\multirow{2}*{NGC5548} & \multirow{2}*{60002044005} & 0.985$\pm$0.006& 3.56$\pm$0.52& 1.74 $\pm$ 0.02 & 1.61e-02 $\pm$ 4.69e-03 & 38 $\pm$ 26 & 20 $\pm$ 5 & \\ 
 & & 0.00 $\pm$ -1.00 & 0.05$^u$& 0.96(520/540) & -10.222 $\pm$ 0.127 & 43.601 $\pm$ 0.127 \\
\multirow{2}*{NGC5548} & \multirow{2}*{60002044006} & 1.020$\pm$0.006& 1.66$\pm$0.23& 1.80 $\pm$ 0.02 & 1.67e-02 $\pm$ 1.46e-03 & 38 $\pm$ 26 & 73 $\pm$ 72 & \\ 
 & & 0.00 $\pm$ -1.00 & 0.05$^u$& 1.07(575/539) & -10.247 $\pm$ 0.040 & 43.576 $\pm$ 0.040 \\
\multirow{2}*{NGC5548} & \multirow{2}*{60002044008} & 0.977$\pm$0.007& 3.91$\pm$0.40& 1.73 $\pm$ 0.03 & 1.24e-02 $\pm$ 9.45e-04 & 38 $\pm$ 26 & 27 $\pm$ 10 & \\ 
 & & 0.00 $\pm$ -1.00 & 0.05$^u$& 0.95(510/535) & -10.329 $\pm$ 0.039 & 43.494 $\pm$ 0.039 \\
\multirow{2}*{NGC5548} & \multirow{2}*{90701601002} & 1.010$\pm$0.008& 1.84$\pm$0.24& 1.78 $\pm$ 0.02 & 1.40e-02 $\pm$ 1.37e-03 & 38 $\pm$ 26 & 84 $\pm$ -1 & \\ 
 & & 0.00 $\pm$ -1.00 & 0.05$^u$& 0.94(447/474) & -10.310 $\pm$ 0.045 & 43.513 $\pm$ 0.045 \\ \hline 
\multirow{2}*{NGC7582} & \multirow{2}*{60061318002} & 0.967$\pm$0.021& 19.25$\pm$3.24& 1.40 $\pm$ 0.06 & 2.49e-03 $\pm$ 2.70e-04 & 60 $\pm$ 13 & 84 $\pm$ -1 & \\ 
 & & 0.30 $\pm$ 0.07 & 0.05$^u$& 1.25(188/151) & -10.789 $\pm$ 0.063 & 42.005 $\pm$ 0.063 \\
\multirow{2}*{NGC7582} & \multirow{2}*{60061318004} & 0.981$\pm$0.026& 41.95$\pm$5.67& 1.47 $\pm$ 0.09 & 3.13e-03 $\pm$ 8.11e-04 & 60 $\pm$ 13 & 84 $\pm$ -1 & \\ 
 & & 0.30 $\pm$ 0.07 & 0.05$^u$& 1.40(158/113) & -10.739 $\pm$ 0.129 & 42.055 $\pm$ 0.129 \\
\multirow{2}*{NGC7582} & \multirow{2}*{60201003002} & 0.990$\pm$0.009& 24.15$\pm$1.38& 1.61 $\pm$ 0.04 & 8.79e-03 $\pm$ 8.23e-04 & 60 $\pm$ 13 & 23 $\pm$ 7 & \\ 
 & & 0.30 $\pm$ 0.07 & 0.05$\pm$0.03& 1.09(537/491) & -10.388 $\pm$ 0.049 & 42.406 $\pm$ 0.049 \\
 \hline 
 \end{tabular}
\end{table*}

\begin{table*}
\caption{Part of \nustar\, data analysis results for the five CLAGNs from the literature. The table lists the source name, observational ID, instrument, the model to fit the spectra, column density from the line of sight, the photon index, X-ray luminosity, and references.}
\centering
\scriptsize
\begin{tabular}{lccc cccc} \hline\hline
Name & ObsID & Instrument   & model  &    $N_{\rm H,los}$  & $\Gamma$ &log$L_{\rm X-ray}$  &Reference   \\  \hline 
ESO362-G18 & 0790810101/60201046002 & XMM-Newton/NuSTAR & double warm corona & 0.66$\pm$0.10 & 1.66$\pm$0.03& 42.59 & \citet{2022RAA....22c5002Z} \\
NGC1365 & 0692840301/60002046005 & XMM-Newton/NuSTAR &  pha*gabs1*gabs2(pow+laor)) & 7.54 $^{+0.5}_{-0} $ & 2.09$\pm$0.05& 42.40 & \citet{2021RAA....21..199L} \\
NGC1365 & 0692840401/60002046007 & XMM-Newton/NuSTAR &  pha*gabs1*gabs2(pow+laor)) & 4.16 $^{+0.4}_{-0} $ & 2.10$\pm$0.05& 42.35 & \citet{2021RAA....21..199L} \\
NGC1365 & 0692840501/60002046009 & XMM-Newton/NuSTAR &  pha*gabs1*gabs2(pow+laor)) & 12.04 $^{+0.6}_{-1} $ & 1.92$\pm$0.06& 42.28 & \citet{2021RAA....21..199L} \\
NGC5548 & 60002044006 & NuSTAR & tbabs*ztbabs(pexrav+zgauss) & 4.46 $^{+2.1}_{-2} $ & 1.79$\pm$0.09& 43.49 & \citet{Pal2022} \\
NGC5548 & 60002044005 & NuSTAR & tbabs*ztbabs(pexrav+zgauss) & 5.61 $^{+2.3}_{-2} $ & 1.62$\pm$0.10& 43.44 & \citet{Pal2022} \\
NGC5548 & 60002044003 & NuSTAR & tbabs*ztbabs(pexrav+zgauss) & 4.77 $^{+1.3}_{-2} $ & 1.81$\pm$0.08& 43.37 & \citet{Pal2022} \\
NGC5548 & 60002044002 & NuSTAR & tbabs*ztbabs(pexrav+zgauss) & 6.06 $^{+1.6}_{-2} $ & 1.75$\pm$0.10& 43.43 & \citet{Pal2022} \\
NGC5548 & 60002044008 & NuSTAR & tbabs*ztbabs(pexrav+zgauss) & 6.91 $^{+2.5}_{-2} $ & 1.61$\pm$0.11& 43.35 & \citet{Pal2022} \\
NGC5548 & 90701601002 & NuSTAR & tbabs*ztbabs(pexrav+zgauss) & 3.66 $^{+1.2}_{-2} $ & 1.81$\pm$0.09& 43.36 & \citet{Pal2022} \\
NGC7582 & 60201003002 & NuSTAR & tbabs(cutoffpl*TBpcf+xillver) & 44.4 $^{+1.5}_{-2} $ & 1.55$\pm$0.03& 41.99 & \citet{2023MNRAS.522.1169L} \\
NGC7582 & 60201003002 & NuSTAR & tbabs(cutoffpl*TBpcf+xillver) & 42.89$\pm$1.50 & 1.65$\pm$0.03& 42.14 & \citet{2023MNRAS.522.1169L} \\
NGC7582 & 60201003002 & NuSTAR & tbabs(cutoffpl*TBpcf+xillver) & 35.88$\pm$2.20 & 1.57$\pm$0.05& 42.10 & \citet{2023MNRAS.522.1169L} \\
NGC7582 & 60201003002 & NuSTAR & tbabs(cutoffpl*TBpcf+xillver) & 50.61 $^{+1.6}_{-2} $ & 1.62$\pm$0.04& 42.15 & \citet{2023MNRAS.522.1169L} \\
NGC7582 & 60201003002 & NuSTAR & tbabs(cutoffpl*TBpcf+xillver) & 55.11 $^{+3.0}_{-3} $ & 1.57$\pm$0.05& 41.87 & \citet{2023MNRAS.522.1169L} \\
NGC7582 & 60061318004 & NuSTAR & tbabs(cutoffpl*TBpcf+xillver) & 62.82 $^{+11.2}_{-9} $ & 1.33$\pm$0.06& 41.54 & \citet{2023MNRAS.522.1169L} \\
NGC7582 & 60061318002 & NuSTAR & tbabs(cutoffpl*TBpcf+xillver) & 32.36 $^{+4.6}_{-4} $ & 1.38$\pm$0.05& 41.63 & \citet{2023MNRAS.522.1169L} \\
NGC7582 & 60201003002 & NuSTAR & tbabs(cutoffpl*TBpcf+xillver) & 43.57 $^{+2.0}_{-2} $ & 1.50$\pm$0.04& 41.90 & \citet{2023MNRAS.522.1169L} \\
NGC7582 & 60201003002 & NuSTAR & tbabs(cutoffpl*TBpcf+xillver) & 45.84$\pm$1.60 & 1.51$\pm$0.04& 42.02 & \citet{2023MNRAS.522.1169L} \\
\hline
\end{tabular}
\label{table:xray}
\end{table*}



\bsp	
\label{lastpage}
\end{document}